\documentclass{aa}
\usepackage{graphicx}
\usepackage{color}
\usepackage{amssymb}
\definecolor{Gray}{cmyk}{0,0,0,0.50}
\usepackage{pifont}

\usepackage{savefnmark}
\usepackage{subfig}
\usepackage{float}
\usepackage{txfonts}
\usepackage{pdflscape}
\newcommand{\cii}{[C\,{\sc ii}]\,}
\begin{document} 

\def\arcsec{\hbox{$^{\prime\prime}$}}

   \title{A spectral stacking analysis to search for faint outflow signatures in $z\sim6$ quasars}

   \titlerunning{Outflows in $z\sim6$ quasars}
   \authorrunning{F.\ Stanley et al.}

   \author{F.\ Stanley, J.~B.~Jolly, S.\ K\"onig \and K.\ K.\ Knudsen
          }

   \institute{Department of Space, Earth and Environment, Chalmers University of Technology, Onsala Space Observatory, SE-439 92 Onsala, Sweden\\
              \email{flora.stanley@chalmers.se}
             }

   \date{Accepted August 2019}

 
  \abstract{}
{Outflows in quasars during the early epochs of galaxy evolution are an important part of the feedback mechanisms potentially affecting the evolution of the host galaxy. However, systematic observations of outflows are only now becoming possible with the advent of sensitive mm telescopes. In this study we use spectral stacking methods to search for faint high velocity outflow signal in a sample of \cii detected, $z\sim6$ quasars.} 
{We search for broad emission line signatures from high-velocity outflows for a sample of 26 $z\sim6$ quasars observed with the Atacama Large Millimeter Array (ALMA), with a detection of the \cii line.
The observed emission lines of the sources are dominated by the host galaxy, and outflow emission is not detected for the individual sources. We use a spectral line stacking analysis developed for interferometric data to search for outflow emission. We stack both extracted spectra and the full spectral cubes. We also investigate the possibility that only a sub-set of our sample contributes to the stacked outflow emission.}
{We find only a tentative detection of a broad emission line component in the stacked spectra. When taking a region of about 2\,\arcsec around the source central position of the stacked cubes, the stacked line shows an excess emission due to a broad component of 1.1--1.5$\sigma$, but the significance drops to 0.4--0.7$\sigma$ when stacking the extracted spectra from a smaller region. The broad component can be characterised by a line width of full width half-max $\rm FWHM > 700$\,km\,s$^{\rm -1}$. Furthermore, we find a sub-sample of 12 sources the stack of which maximises the broad component emission.The stack of this sub-sample shows an excess emission due to a broad component of 1.2--2.5$\sigma$. The stacked line of these sources has a broad component of $\rm FWHM > 775$\,km\,s$^{\rm -1}$.}
{We find evidence suggesting the presence of outflows in a sub-sample of 12 out of 26 sources, and have demonstrated the importance of spectral stacking techniques in tracing faint signal in galaxy samples. However, deeper ALMA observations are necessary to confirm the presence of a broad component in the individual spectra.}
 

   \maketitle
%

\section{Introduction}
Our understanding of redshift $z$\,$\gtrsim$\,6 galaxies and super-massive black holes (SMBHs) has increased significantly, and still continues to do so. More and more quasars are detected at high redshifts, with their numbers now in the hundreds at redshifts $z$\,$>$5.5 \citep[e.g.,][]{Banados16,Jiang16}. These quasars are known to have BH masses of $>$\,$10^9$\,$M_\odot$ \citep[e.g.,][]{Fan06,Kurk07,DeRosa11}. Their hosts tend to be massive galaxies with stellar masses of 10$^{10}-10^{11}$\,M$_\odot$. Star formation rates (SFRs) in $z$\,$\sim$\,6 quasar host galaxies can range between 10--3000\,M$_\odot$\,yr$^{-1}$ \citep[e.g.,][]{Omont13, Willott13, Venemans12, Venemans16, Wang16, willott17, venemans18}, and they appear gas and dust-rich \citep[e.g.,][]{Maiolino05,Wang11,Wang13,Venemans17a}. However, we are still lacking understanding of if and how quasar feedback affects the host galaxy and the surrounding environment. \\
\indent
Massive outflows observed in active galaxies are linked with the feedback that regulates the overall star formation and BH growth \citep[e.g.,][]{fabian12,king15}. Observational studies of local and low-redshift galaxies have yielded a large number of outflow detections, both in the ionized and the molecular gas phases, that are seen in starburst galaxies as well as active galactic nuclei \citep[AGN, e.g.,][see also review by \citealt{fabian12}]{rupke11,sturm11,Maiolino12,Veilleux13,Cicone14,harrison14,Tadhunter18,barcosmunoz18,Gallerani18}.
Furthermore, simulations of galaxy formation and evolution find that feedback is necessary in order to reproduce the observed galaxy stellar mass functions \citep[e.g.,][]{somerville08,sijacki15,Schaye15}. 
Therefore, it is necessary to characterise the physical properties of quasar outflows observationally, as well as from simulations, to constrain their feedback mechanisms.\\
\indent
As the cosmic SFR peaks only a few billion years after the Big Bang, the detection and characterisation of outflows in the high redshift quasar populations is a necessity if we are to understand early galaxy evolution. For high-redshift galaxies, the number of detections is still limited, in part because of the large distances resulting in the faintness of the sources, and the limited spatial resolution. Currently, only a single detection of a [CII] outflow is known in a quasar at redshift $z\sim6$ - \object{SDSS J1148+5251} \citep[e.g.,][]{Maiolino12,Cicone15}. However, recent mm observations of $z>5$ quasars \citep[e.g.][]{Venemans16,Venemans17a,Venemans17b,willott17,Decarli17,Decarli18} and star-forming galaxy samples \citep[e.g.][]{Gallerani18} have allowed for the search of outflow signatures in more sources. In a study of 27 quasars at $z\sim6$, \cite{Decarli18} found a high detection rate of 85\% in \cii, in contrast to normal star-forming galaxies \citep[e.g.,][]{Knudsen16}. Finding no evidence of outflows in the individual sources, \cite{Decarli18} examined the possibility of an outflow signature in the stacked spectra of the full sample. The stacked spectra showed no evidence of an outflow component; however, the authors only took into account the signal within the central pixel of each target. Interestingly, a study of 9 star-forming galaxies at similar redshifts of $z\sim5.5$ observed in \cii shows excess emission in their stacked spectra at high velocities ($\sim1000$\,km/s) that could be attributed to the presence of outflows \citep{Gallerani18}.   \\
\indent
In this study we aim to use recent ALMA archival data to examine the presence of outflows in a sample of $z\sim6$ quasars observed in \cii. Using extensive spectral stacking analysis over the extended galaxy regions, we examine the presence of an outflow component in the stacked \cii line, over a larger area. Furthermore, we use sub-sampling to determine a sub-sample of sources most likely to have an outflow component when stacked. We note that this paper was written in parallel to the recent article by \cite{Bischetti18} with a similar aim to our work. However, our sample selection, redshift range, and analysis are different. \\
\indent In Sect.~\ref{sec:sample} we present the sample of quasars selected from the ALMA archive along with a description of the projects they come from. In Sect.~\ref{sect:observations} we describe the data reduction and analysis. 
In Sect.~\ref{sect:stackingmethods} we present methods used in our stacking analysis. The results are given in Sect.~\ref{sect:results}, followed by the discussion in Sect.~\ref{sec:disc}. Finally, in Sect.~\ref{sec:summary} we give a brief summary and the conclusions of this work. Throughout this paper we assume $ H_0 = 70 \rm km \, s^{-1} \, Mpc^{-1}$,
$\rm \Omega_M = 0.3$, $\rm \Omega_\Lambda = 0.7$.

\section{Sample}\label{sec:sample}
\begin{figure}[t]
  \begin{center}
  
    \includegraphics[width=0.98\columnwidth]{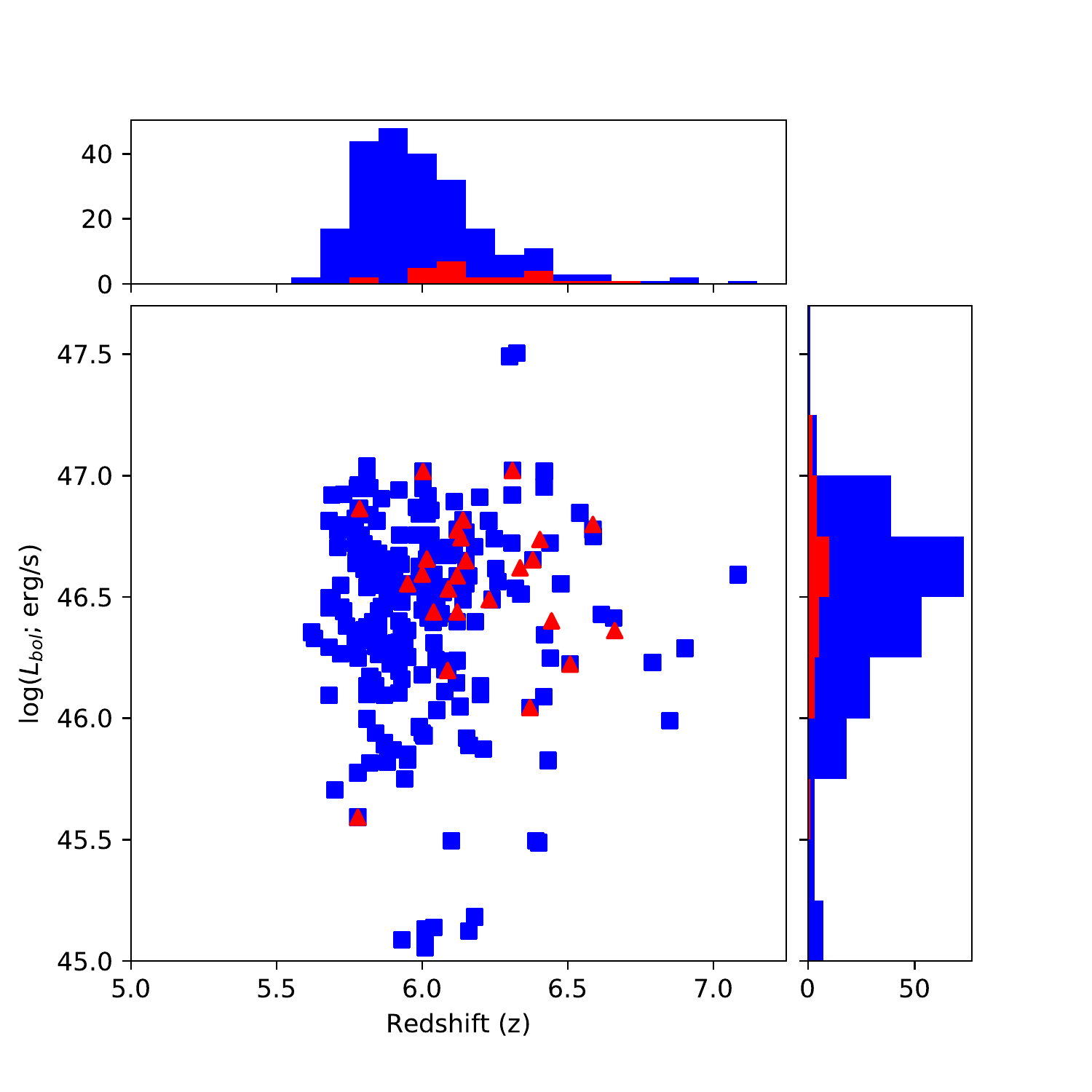}
   \caption{The AGN bolometric luminosity (L$_{bol}$) as a
function of redshift ($z$) of our sample (red triangles). Also plotted are 
the known high redshift quasars from the compiled catalogue of \citet{Banados16}, 
the catalogue of SDSS detections of \citet{Jiang16} and the new detections of \citet{Mazzucchelli17},
from which the majority of our sample has been selected.}\label{fig:sample}
   	
    \end{center}   
\end{figure}

To study the outflow activity in distant quasars, we selected sources at $z\sim6$ that have been observed with ALMA in the far-infrared fine-structure line \cii $^{\rm 2}$P$_{\rm 3/2}$$-$$^{\rm 2}$P$_{\rm 1/2}$, at a rest frequency of 1900.537~GHz ($\lambda$\,$\sim$158~$\mu$m). The sample was selected from the ALMA archive from relatively recent cycles (observing cycle~3 and above) to ensure that all sources where observed with similar technical specifications, such as the number of antennas and $uv$-coverage. Additionally, the sample was limited to the sources for which the data was public by January 2018.\\
\indent
Our sample consists of 32 optically-selected quasars (see Table~\ref{tab:sample_overview}) from projects \textit{2015.1.00606.S} \citep{willott17}, \textit{2015.1.00997.S} (PI: R. Maiolino), and \textit{2015.1.01115.S} \citep{Decarli18}. The sample has redshifts of $5.779 < z < 6.661$, and absolute magnitudes at 1450$\AA$ of $-27.8 < M_{\rm 1450\AA} < -23.89$ that correspond to bolometric luminosities of $3.9\times10^{45}<L_{\rm bol}<10^{47}$ erg\,s$^{-1}$ \citep[converted following][see table~1]{Runnoe12,Venemans16}. In Fig.~\ref{fig:sample} we show the distribution of the $L_{\rm bol}$ together with the distribution of the redshift, for a total of 321 quasars from \cite[][and references therein]{Banados16}, \cite{Jiang16}, and \cite{Mazzucchelli17}. The sample of quasars used in our analysis are overplotted as red triangles. Our sample covers the majority of the bolometric luminosity range of the general detected population, while concentrating on the higher end of the redshift distribution by design. Therefore, it is representative of optically selected quasars at redshifts of $z = 5.8-6.7$. \\
\indent
As the purpose of this study is to determine the presence of an outflow component, we only include the sample sources where the \cii line was detected. As a result, we find that 26 sources are detected, and the data are of sufficient quality to be used for the purpose of this paper. This final sample is provided in Table~\ref{tab:results_cii}.

\begin{table*}
	\begin{center}
        \caption{
        Properties for our sample of $z\sim6$ quasars available in the ALMA archive by January 2018.  (a) The identification (ID) of the source, with the associated numbers corresponding to the references of the source \citep[including][in which the ALMA data have been published]{Bischetti18,Decarli18,willott17}; (b) \& (c) the J2000 RA \& DEC coordinates of the sources from the optical; (d) redshift of source in the literature; (e) the bolometric AGN luminosity estimated from the 1450$\AA$ absolute magnitudes found in the literature; (f) The ProjectID of the ALMA project the source was observed in, where A=2015.1.00606.S, B=2015.1.00997.S, C=2015.1.01115.S; (g) flag on wether or not we detect the [C{\sc ii}] line in our analysis, (h) The original beamsize of the natural weighted images, before re-imaging to a 0.8\arcsec\,$\times$\,0.8\arcsec beamsize.
       }\label{tab:sample_overview}
		\begin{tabular}{lcccccccc}
			\hline
			 ID$^{(a)}$ & RA$^{(b)}$	& DEC$^{(c)}$	& $z^{(d)}$ & $\log{L_{\rm bol}}^{(e)}$ & Project$^{(f)}$ & [C{\sc ii}]$^{(g)}$  & Original $^{(h)}$  \\
			 	         & hh:mm:ss & dd:mm:ss &    	&		   			&          & detected  & beamsize \\
			\hline
			\hline
            \noalign{\smallskip}
			PSO J007.0273+04.9571$^{1,2,6}$	& 00:28:06.56	& $+$04:57:25.7	& 6.00 & 46.59 & C & Y & 0.65\arcsec\,$\times$\,0.47\arcsec \\
			PSO J009.7355-10.4316$^{2,6}$	& 00:38:56.52	& $-$10:25:53.9	& 5.95 & 46.55 & C &	Y & 0.63\arcsec\,$\times$\,0.44\arcsec \\
			SDSS J0129-0035$^{3,16}$  & 01:29:58.51	& $-$00:35:39.7 & 5.779 & 45.59  &	 B	& Y & 0.45\arcsec\,$\times$\,0.38\arcsec \\
			VST-ATLAS J025.6821–33.4627$^{2,4,6}$  & 01:42:43.73	& $-$33:27:45.5	& 6.31 & 47.02	& C & Y & 0.87\arcsec\,$\times$\,0.75\arcsec \\
			(J0142-3327) & & & & & & & \\
			PSO J065.4085-26.9543$^{2,6}$ & 04:21:38.05  & $-$26:57:15.6	& 6.14 & 46.82 & C &	Y & 1.11\arcsec\,$\times$\,0.83\arcsec \\
			PSO J065.5041-19.4579$^{2,6}$ & 04:22:00.99	& $-$19:27:28.7	& 6.12 & 46.59 & C &	Y & 1.09\arcsec\,$\times$\,0.75\arcsec \\
			VDESJ0454-4448$^{2,6,13}$  & 04:54:01.79  & $-$44:48:31.1 & 6.09 & 46.53 & C &	Y & 1.12\arcsec\,$\times$\,0.79\arcsec \\
			SDSS J0842+1218$^{2,5,6,8}$	& 08:42:29.23  & $+$12:18:48.2   & 6.069 & 46.69 & C & N$^*$ & 1.26\arcsec\,$\times$\,1.12\arcsec \\
			PSO J159.2257-02.5438$^{2,6}$ & 10:36:54.19  & $-$02:32:37.9   & 6.38  & 46.65 & C & Y & 1.28\arcsec\,$\times$\,0.98\arcsec \\
			SDSS J1030+0524$^{2,6}$	& 10:30:27.10  & $+$05:24:55.0   & 6.308 & 46.72 & C & N & 1.21\arcsec\,$\times$\,0.98\arcsec \\
			SDSS J1044-0125$^{2,3}$  & 10:44:33.04  & $-$01:25:02.2	& 5.7847 & 46.86 & B & Y & 0.74\arcsec\,$\times$\,0.70\arcsec \\
			VIK J1048-0109$^{6,15}$	 & 10:48:19.09  & $-$01:09:40.3 	& 6.661	& 46.36 & C & Y & 1.46\arcsec\,$\times$\,1.00\arcsec \\
			PSO J167.6415-13.4960$^{2,6,14,19}$ & 11:10:33.98  & $-$13:29:45.6  & 6.508 & 46.22 & A,C  & Y & 0.77\arcsec\,$\times$\,0.63\arcsec \\  
			ULAS J1148+0702$^{2,6}$	& 11:48:03.29  & $+$07:02:08.3   & 6.32  & 46.54 & C & N & 1.32\arcsec\,$\times$\,1.08\arcsec \\
			VIK J1152+0055$^{2,6,10}$ & 11:52:21.27  & $+$00:55:36.7   & 6.37  & 46.04 & C & Y & 1.26\arcsec\,$\times$\,1.02\arcsec \\			
			ULAS J1207+0630$^{2,6}$  & 12:07:37.44  & $+$06:30:10.4   & 6.04  & 46.59 & C & N$^*$ & 1.73\arcsec\,$\times$\,0.89\arcsec \\
			PSO J183.1124+05.0926$^{2,6}$	& 12:12:26.98  & $+$05:05:33.5   & 6.4039 & 	46.73 & C & Y & 1.25\arcsec\,$\times$\,1.05\arcsec \\
			SDSS J1306+0356$^{2,6,7}$	 & 13:06:08.26  & $+$03:56:26.3   & 6.016 & 46.66 & C & Y & 1.11\arcsec\,$\times$\,0.90\arcsec \\
			ULAS J1319+0950$^{2,6,12}$  & 13:19:11.29  & $+$09:50:51.4   & 6.133 & 46.74 &  B & Y & 1.20\arcsec\,$\times$\,1.04\arcsec \\
			PSO J217.0891-16.0453$^{2,6}$ & 14:28:21.39  & $-$16:02:43.3   & 6.11  & 46.70 & C & N$^*$ & 1.17\arcsec\,$\times$\,0.86\arcsec \\
			CFHQS J1509-1749$^{2,6,17}$	& 15:09:41.78  & $-$17:49:26.8   & 6.121 & 46.78 & C & Y & 1.36\arcsec\,$\times$\,0.87\arcsec \\ 
			PSO J231.6576-20.83335$^{5,6,11}$	& 15:26:37.84  & $-$20:50:00.7 & 6.586  & 46.80 & C & Y & 1.21\arcsec\,$\times$\,0.88\arcsec \\
			PSO J308.0416-21.2339$^{2,5,6}$  & 20:32:09.99  & $-$21:14:02.3   & 6.23 & 46.49 & C & Y & 0.85\arcsec\,$\times$\,0.65\arcsec \\
			SDSS J2054-0005$^{2,3}$  & 20:54:06.49  & $-$00:05:06.49  & 6.0391 & 46.44 &  B & Y	& 0.69\arcsec\,$\times$\,0.67\arcsec \\
			CFHQS J2100-1715$^{2,5,6,18}$ & 21:00:54.62  & $-$17:15:22.5 & 6.087  & 46.20 & C & Y & 0.73\arcsec\,$\times$\,0.63\arcsec \\
			VIK J2211-3206$^{6,15}$	 & 22:11:12.39  & $-$32:06:12.9 &  6.336 & 46.62 & C & Y & 0.84\arcsec\,$\times$\,0.67\arcsec \\
			VIMOS2911001793$^{6,19}$ & 22:19:17.22 & $+$01:02:48.90 & 6.16 & (-22.60) & A & Y & 0.77\arcsec\,$\times$\,0.68\arcsec \\ 
			PSO J340.2041-18.6621$^{2,6}$ & 22:40:49.00  & $-$18:39:43.8	& 6.01   & 46.51 & C & N & 0.78\arcsec\,$\times$\ 0.68\arcsec \\
			SDSS J2310+1855$^{2,3,9}$  & 23:10:38.88  & $+$18:55:19.7   & 6.0031 & 47.02 & B & Y & 1.10\arcsec\,$\times$\,0.73\arcsec \\
			VIK J2318-3113$^{15}$  &  23:18:18.35  & $-$31:13:46.4   & 6.444 & 46.40 & C & Y & 0.82\arcsec\,$\times$\,0.78\arcsec \\
			VIK J2318-3029$^{15}$  & 23:18:33.10  & $-$30:29:33.4   & 6.12  & 46.44 & C & Y & 0.91\arcsec\,$\times$\,0.74\arcsec \\
			PSO J359.1352-063831$^{2,6}$ & 23:56:32.00     & $-$06:22:59.0     & 6.15 & 46.65 & C & Y & 1.06\arcsec\,$\times$\,0.62\arcsec \\
\hline
\noalign{\smallskip}
		\end{tabular}
		\end{center}
		\footnotesize{References: 1: \cite{Banados14}; 2: \cite{Banados16}; 3: \cite{Bischetti18}; 4: \cite{Carnall15};
		 5: \cite{Decarli17}; 6: \cite{Decarli18}; 7: \cite{Fan01}; 8: \cite{Jiang15}; 9: \cite{Jiang16};
		 10: \cite{Matsuoka16}; 11: \cite{Mazzucchelli17}; 12: \cite{Mortlock09}; 13: \cite{Reed15}; 
		 14: \cite{Venemans15}; 15: Venemans et al. (in prep.); 16: \cite{Wang13}; 17: \cite{Willott07}; 
		 18: \cite{Willott10}; 19: \cite{willott17}. $^*$These sources are not detected in our analysis; however if only the central pixel of the source is taken the \cii line is detected in agreement with \cite{Decarli18}.}
	
\end{table*}

\section{ALMA data reduction and analysis}
\label{sect:observations}
After extracting the \cii raw data of the 31 quasars from the ALMA archive, we calibrated and imaged (including continuum subtraction) each data set with the Common Astronomy Software Applications\footnote{http://casa.nrao.edu/ \citep[CASA,][]{McMullin07}}. For about 70\% of the sources, a careful manual calibration of the observations was necessary to warrant that the data were calibrated correctly. This was necessary in order to address issues found with flux calibration\footnote{This is the case for sources originally calibrated with CASA version 4.6 or prior with the sources Ceres or Pallas as their flux calibrators. Through private communication with the ALMA Nordic ARCnode it was pointed out to us that for those versions of CASA the flux models for Pallas and Ceres were wrong. Consequently, the the fluxes extrapolated for the sources observed with Ceres or Pallas as flux calibrators will be wrong without re-calibration.}, incorrect antenna positions\footnote{In these cases the data were observed right after a shift in the antenna positions but did not have the new correct coordinates when original calibration was performed. We have made sure to use the correct coordinates.}, low source amplitudes requiring additional flagging, and other issues found in the original calibration. Specifically for the sources with flux calibration issues we found that the fluxes where wrong by $\sim$20--45\%. For the remaining $\sim$30\% of the sources, the calibration results from the ALMA pipeline were sufficient. For the manually calibrated data sets, we used the CASA software version 5.1.1, to ensure that bugs and calibration issues found for previous CASA versions have been corrected for. Hereafter, a few details are given for each of the archival ALMA projects:\\
\indent
\textit{2015.1.00606.S:} As part of the work presented here, two sources of this program were extracted from the ALMA archive. Observations took place on March 22, 2016, and April 27, 2016. 37 antennas were included in the array with minimum and maximum baselines of 15.3~m and 460.0~m, respectively. Applying natural weighting, the resulting beam size is 0.8\arcsec\,$\times$\,0.6\arcsec. The spectral setup consisted of four spectral windows of 2.0~GHz bandwidth, each containing 128 channels of 15.625~MHz width. \\
\indent
\textit{2015.1.00997.S:} The five sources (see Table\,\ref{tab:sample_overview}) we included from this program have been observed between January and July 2016. Between 37 and 49 antennas were included in the arrays with baselines between 15.1~m and 1.0~km. The synthesised beam sizes resulting from natural weighting ranged between $\sim$0.4\arcsec\ and $\sim$1.2\arcsec. The observations were setup with four spectral windows of 1.875~GHz bandwidth each, 480 channels per spectral window, and 3.9~MHz wide channels.\\
\indent
\textit{2015.1.01115.S:} The 26 sources taken from this program were included in our sample. The data were obtained between January and June 2016 in arrays with baselines between 15.1~m and 704.1~m. Natural weighting led to synthesised beam sizes between $\sim$0.4\arcsec\ and 1.7\arcsec. For the spectral setup, 960 channels of 1.953~MHz width covered each of the four 1.875~GHz wide spectral windows. For more extensive details see \citet{Decarli18}.\\

\begin{figure}[t]
  \centering
    \includegraphics[width=0.48\textwidth]{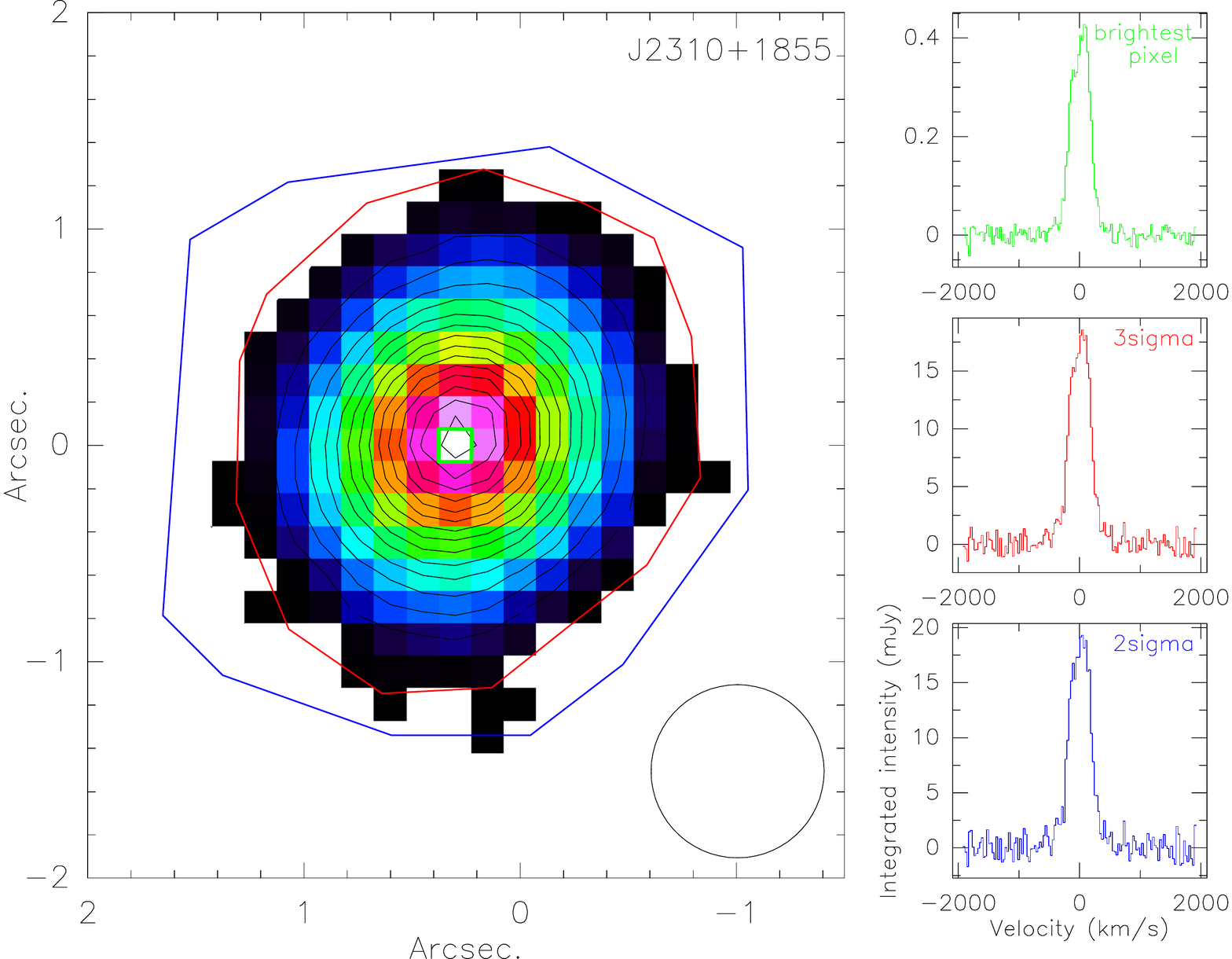}
   \caption{Example for an integrated intensity distribution (moment zero map) and spectra 
    extracted from different regions in \object{J2310+1855}. The green box marks the area of the brightest pixel, the area marked by the red line contains all signal above the 3$\sigma$ threshold, and the blue line outlines the region with signals above 2$\sigma$. The spectra extracted from these three regions are shown on the right-hand side. They are color-coded according to the region from which they were extracted, and the region of extraction is mentioned in the panel for each spectrum as well. On the lower-left side of the image we also show the beam size of our images.}
   \label{fig:extract_spectra_example}
\end{figure}

Finally, all data cubes were imaged to the same angular resolution of $0.8''\times0.8''$ and spectral resolution of 30~km\,s$^{\rm -1}$, with a pixel scale of 0.2\arcsec. Continuum subtraction was performed using the CASA task {\sc uvcontsub}, using the channels with no line emission. 
The uncertainty of the absolute flux calibration for all sources falls within 5--10\%.
For further spectral analysis, the resulting data cubes were imported into the MAPPING software package within GILDAS\footnote{http://www.iram.fr/IRAMFR/GILDAS}. From the continuum-subtracted cubes, we created integrated intensity maps (moment-0) of the \cii line for each source. With the help of these maps we extracted 1-D spectra from the position of the brightest pixel, and the individual areas enclosed by the 2 and 3$\sigma$ integrated intensity contours (for an example see Fig.\,\ref{fig:extract_spectra_example}). We extracted the continuum flux densities ($S_{\rm cont}$) of each source using a 2D Gaussian fit to the continuum source. The IR (8--1000$\mu m$) luminosity of the sources was estimated by normalising a set of star-forming galaxy templates (from \citealt{Mullaney11}, see also \citealt{Stanley18}) to the measured $S_{\rm cont}$ values. 

\begin{table*}
       \begin{center}
       \caption{
       Results of the observations for the sample sources. We note that only sources with a \cii detection are included here (see Table~\ref{tab:sample_overview}). (a) Source ID; (b) Observed \cii frequency; (c) \cii redshift calculated from the observed line; (d) the rms of the spectrum from the final imaging with beamsize of 0.8\arcsec$\times$0.8\arcsec, and channelwidths of 30\,km/s; (e) the peak flux density of the \cii line; (f) the band-6 continuum flux density; (g) the IR (8--1000$\mu$m) luminosity calculated from the band-6 continuum. In bold we highlight the sources included in the {\em max sub-sample}.
        }\label{tab:results_cii}
	
		\begin{tabular}{lcccccccc}
			\hline 
			 ID$^{(a)}$ & [CII] freq.$^{(b)}$ & z$_{\rm [CII]}$$^{(c)}$ & rms$^{(d)}$   & $S_{\rm peak}^{(e)}$ & FWHM &$S_{\rm cont}^{(f)}$  & $L_{\rm IR}^{(g)}$  \\
			 			& [GHz]	      & 			    & [mJy/beam] & [mJy] & [km/s] & [mJy] & [$10^{12} L_\odot$] \\
			 \hline
             \hline
             \noalign{\smallskip}
			 PSO J007.0273+04.9571  & 271.463 & 6.001 & 0.39 & 4.89 & 965 & $1.87 \pm 0.04$ & $2.09 \pm 0.48$ \\
			 PSO J009.7355-10.4316  & 271.364 & 6.004 & 0.36 & 17.32 & 583 & $1.60 \pm 0.04$ & $1.78 \pm 0.41$ & \\
			 SDSS J0129-0035 & 280.390 & 5.778 & 0.10 & 10.60 & 384 & $2.65 \pm 0.02$ & $2.9 \pm 0.7$ \\
			 VST-ATLAS J025.6821–33.4627 & 259.010 & 6.338 & 0.70 & 12.16 & 656 & $1.19 \pm 0.05$ & $1.4 \pm 0.3$ \\
			 {\bf PSO J065.4085-26.9543}  & 264.429 & 6.187 & 0.79 & 7.12 & 854 & $ 1.00 \pm 0.05 $ & $1.2 \pm 0.3$\\
			 {\bf PSO J065.5041-19.4579} & 266.753 & 6.125 & 0.73 & 3.47 & 777 & $ 0.38 \pm 0.05 $ & $0.4 \pm 0.1$ \\
			 VDESJ0454-4448 & 269.272 & 6.058 & 0.62 & 1.68 & 726 & $ 0.55 \pm 0.04$ & $0.6 \pm 0.1$ \\
			 PSO J159.2257-02.5438  & 257.450 & 6.382 & 0.54 & 3.90 & 717 & $ 0.53 \pm 0.04$ & $0.6 \pm 0.1$ \\
			 SDSS J1044-0125 & 280.106 & 5.786 & 0.24 & 4.16 & 895 &  $ 3.03 \pm 0.02$ & $3.3 \pm 0.8$\\
			 VIK J1048-0109 & 247.600 & 6.676 & 0.49 & 7.42 & 707 & $2.50 \pm 0.04$ & $3.1 \pm 0.7$\\
			 PSO J167.6415-13.4960   & 252.932 & 6.514 & 0.21 & 7.97 & 1003 & $0.58 \pm 0.04$ & $0.7 \pm 0.2$\\
			 {\bf VIK J1152+0055} & 258.045 & 6.365 & 0.74 & 2.27 & 386 & $0.09 \pm 0.03$ & $0.1\pm0.04$ \\ 
			 {\bf PSO J183.1124+05.0926}   & 255.500 & 6.438 & 0.66 & 19.05 & 763 & $3.99 \pm 0.05$ & $5 \pm 1$\\
			 SDSS J1306+0356 & 270.181 & 6.034 & 0.76 & 9.46 & 598 &  $0.82 \pm 0.04$ & $0.9 \pm 0.2$\\
			 ULAS J1319+0950 & 266.431 & 6.133 & 0.37 & 9.05 & 1016 & $4.59 \pm 0.03$ & $5 \pm 1$\\
			 {\bf CFQS J1509-1749} & 266.859 & 6.122 & 0.66 & 3.54 & 1135 & $ 1.46 \pm 0.04$ & $1.7 \pm 0.4$\\
			 {\bf PSO J231.6576-20.83335}  & 250.490 & 6.587 & 0.68 & 8.13 & 922 & $3.34 \pm  0.12$ & $4 \pm 1$\\
			 {\bf PSO J308.0416-21.2339}  & 262.616 & 6.237 & 0.56 & 6.20 & 1128 & $ 0.53 \pm 0.05$ & $6.1 \pm 0.1$\\
			 {\bf SDSS J2054-0005} & 270.007 & 6.039 & 0.33 & 13.81 & 441 & $2.60 \pm 0.03$ & $2.9 \pm 0.7$\\
			 CFQS J2100-1715 & 268.393 & 6.081 & 0.51 & 3.16 & 870 & $0.36 \pm 0.05$ & $0.4 \pm 0.1 $ \\
			 VIK J2211-3206 & 258.963 & 6.339 & 0.57 & 0.71 & 807 & $0.59 \pm 0.03$ & $0.7 \pm 0.2$\\
			 {\bf VIMOS2911001793} & 265.832 & 6.14 & 0.22 & 8.8 & 270 & $0.76 \pm 0.03$ & $ 0.86\pm 0.19 $\\
			 {\bf SDSS J2310+1855} & 271.397 & 6.003 & 0.38 & 19.71 & 788 & $6.97 \pm 0.05$ & $8 \pm 2$\\
			 {\bf VIK J2318-3113} & 255.329 & 6.443 & 0.84 & 2.96 & 609 & $ 0.36 \pm 0.04$ & $0.4 \pm 0.1$\\
			 {\bf VIK J2318-3029} & 265.960 & 6.146 & 0.93 & 7.36 & 582 & $ 2.33 \pm 0.06$ & $2.7 \pm 0.6$\\
             PSO J359.1352-06.3831 & 265.007 & 6.172 & 0.84 & 6.53 & 626 & $0.60 \pm 0.06$ & $0.7 \pm 0.2 $\\
             \hline
             \noalign{\smallskip}
			 \end{tabular}
	\end{center}
\end{table*}



\section{Spectral line stacking techniques}
\label{sect:stackingmethods}

To search for outflow signatures in our sample of $z$\,$\sim$6 quasars, we use the spectral stacking analysis tool {\sc Line Stacker} (Jolly et al. in prep). Stacking the \cii line of the quasars in our sample will allow us to search for a broad emission line component, that is weak in comparison to the bright \cii main line component, and is typically undetected in the individual spectra. We only use the sources where \cii has been detected, as the detection provides the redshift, and the systemic velocity of the line of each quasar host galaxy.\\
\indent

\subsection{Velocity re-binning} \label{subsec:rebin}

To account for the range of line widths (FWHM) observed for the \cii lines in our sample, we choose to normalise all line widths to the same value. The \cii lines of all sources are re-binned in channel space with respect to the smallest FWHM measured in the sample, so that each \cii FWHM is covered by the same number of channels albeit of different channel width. We chose to normalise to the narrowest line and not the widest to avoid oversampling of the data. We note that any sub-samples selected use the re-binned spectra as defined for the full sample. 
For a \cii line that has a larger FWHM than the narrowest line, the channels are re-defined to have a width of: $cw_{rebin}\,=\,cw_{orig}\,\times\,\rm FWHM_{orig}/FWHM_{min}$, where $cw_{rebin}$ is the channel width after re-binning, $cw_{orig}$ is the original channel width of 30\,km\,s$^{\rm -1}$ (see Sect.~\ref{sect:observations}), and FWHM$_{min}$ is the FWHM of the narrowest line, which corresponds to 270\,km/s.\\  
\indent
As a result of the re-binning, the outer velocity bins of the stacked spectrum are not sampled by the same number of sources as the central bins close to the \cii line core. For the analysis we only consider the central part of the spectrum where the channels include information from the majority of the stacked sample. As we are looking at the mean stacked spectra of the sample, the channel width of the stacked spectrum is taken as the mean channel width of all the individual spectra once re-binned, for our sample that corresponds to 60~km\,s$^{\rm -1}$.

\subsection{1-D and 3-D stacking} \label{subsect:cubestacking}
In our analysis we perform both 1-D and 3-D stacking. For the 1-D stacking, we first extract the integrated spectrum for each source, from the region where the moment zero maps have a $>$\,2\,$\sigma$ emission. We then re-bin the spectra as described in Sect.~\ref{subsec:rebin}, and stack. For the 3-D stacking, we take the full spectral cubes of each source. Following the method described in Sect.~\ref{subsec:rebin} we re-bin the full cube using the same parameters determined in our 1-D analysis. The cubes are then stacked pixel by pixel in order to create the stacked cube. The stacked cube allows us to extract the stacked spectrum from different regions as well as image different parts of the stacked spectrum.
 
\subsection{Weighting schemes for the line stacking}
\label{subsec:weighting}

In our stacking analysis we use 3 different types of weights when stacking the spectra: 
\begin{itemize}
    \item[$\bullet$] 
    $w=1$, i.e., no weighting.  
    \item[$\bullet$] 
    $w=1/\sigma_{rms}^2$, i.e., weighting by the noise. In this case, the weight is applied in each velocity channel of the re-binned spectra, using the corresponding noise in that channel. Like this, we take the variations in noise levels of the different observations and spectral channels into account. This is important, as we are interested in the relatively weak outflow component.
    \item[$\bullet$] 
    $w=1/S_{\rm peak}$, i.e., weighting by the peak flux of the detected \cii line ($S_{\rm peak}$). With these weights, we normalise for the varying strength of the main component, so that bright lines do not dominate the result.
\end{itemize}
We repeat our stacking analysis for the three different weighting schemes to search for an outflow component and determine the effects of the varying \cii line strengths, and noise levels.


%
%


\section{Results}
\label{sect:results}
\subsection{Stacking the full sample}
\label{subsec:res_fullsample}


\begin{table*}
	\begin{center}
	\caption{Results of our 1-D stacking analysis. The table includes the peak flux density ($S_{\rm peak}$), FWHM, integrated flux density ($S_{\rm int}$), and central frequency ($\nu_{\rm cen}$) of both the narrow and broad components of the stacked lines.}\label{tab:results_cii_1D}
	\begin{tabular}{lcccccccccc}
\hline 
\hline 
 & \multicolumn{4}{c}{Narrow comp.} & \multicolumn{4}{c}{Broad comp.} \\
 & $S_{\rm peak}$ & FWHM & $S_{\rm int}$ & $\nu_{\rm cen}$ & $S_{\rm peak}$ & FWHM & $S_{\rm int}$ & $\nu_{\rm cen}$ \\ 
 & [mJy] & [km\,s$^{-1}$] & [Jy\,km\,s$^{-1}$] & [GHz] & [mJy] & [km\,s$^{-1}$] & [Jy\,km\,s$^{-1}$] & [GHz] \\ 
\hline 
\underline{Full sample} \\ 
w = 1 & $7.1\pm1.5$ & $357\pm40$ & $2.7\pm0.3$ & $1\pm7$ & $0.7\pm1.5$ & $732\pm596$ & $0.5\pm1.3$ & $-13\pm115$\\ 
w = 1/$\sigma^2$ &  $6.0\pm0.8$ & $368\pm34$ & $2.4\pm0.4$ & $0\pm8$ & $0.3\pm0.8$ & $885\pm996$ & $0.3\pm0.8$ & $3\pm219$ \\ 
w = 1/$S_{peak}$ & $3.7\pm2.3$ & $356\pm98$ & $1.4\pm0.9$ & $0\pm16$ & $0.56\pm2.39$ & $671\pm888$ & $0.4\pm1.8$ & $-7\pm149$ \\ 
\underline{\em max sub-sample} \\ 
w = 1 & $7.9\pm0.6$ & $352\pm24$ & $2.9\pm0.3$ & $2\pm6$ & $0.8\pm0.6$ & $999\pm398$ & $0.8\pm0.7$ &  $-74\pm124$ \\ 
w = 1/$\sigma^2$ & $6.9\pm1.0$ & $335\pm33$ & $2.5\pm0.5$ & $-4\pm7$ & $1.2\pm1.1$ & $745\pm270$ & $1.0\pm1.0$ & $42\pm72$ \\ 
w = 1/$S_{peak}$ & $4.5\pm0.9$ & $311\pm47$ & $1.5\pm0.4$ & $1\pm11$ & $1.2\pm0.9$ & $787\pm277$ & $1.0\pm0.8$ & $-21\pm67$ \\ 
\underline{\em min sub-sample} \\ 
w = 1 & $6.9\pm0.2$ & $375\pm15$ & $2.8\pm0.1$ & $0\pm6$ & .. & .. & .. & .. \\ 
w = 1/$\sigma^2$ &  $5.4\pm0.2$ & $386\pm14$ & $2.2\pm0.1$ & $0\pm6$ & .. & .. & .. & ..\\ 
w = 1/$S_{peak}$ &  $3.5\pm0.2$ & $390\pm26$ & $1.4\pm0.1$ & $0\pm11$ & .. & .. & .. & ..\\ 
\hline  
		\end{tabular}
	\end{center}
\end{table*}

\begin{figure*}[t]
  \begin{center}
      \includegraphics[width=\textwidth]{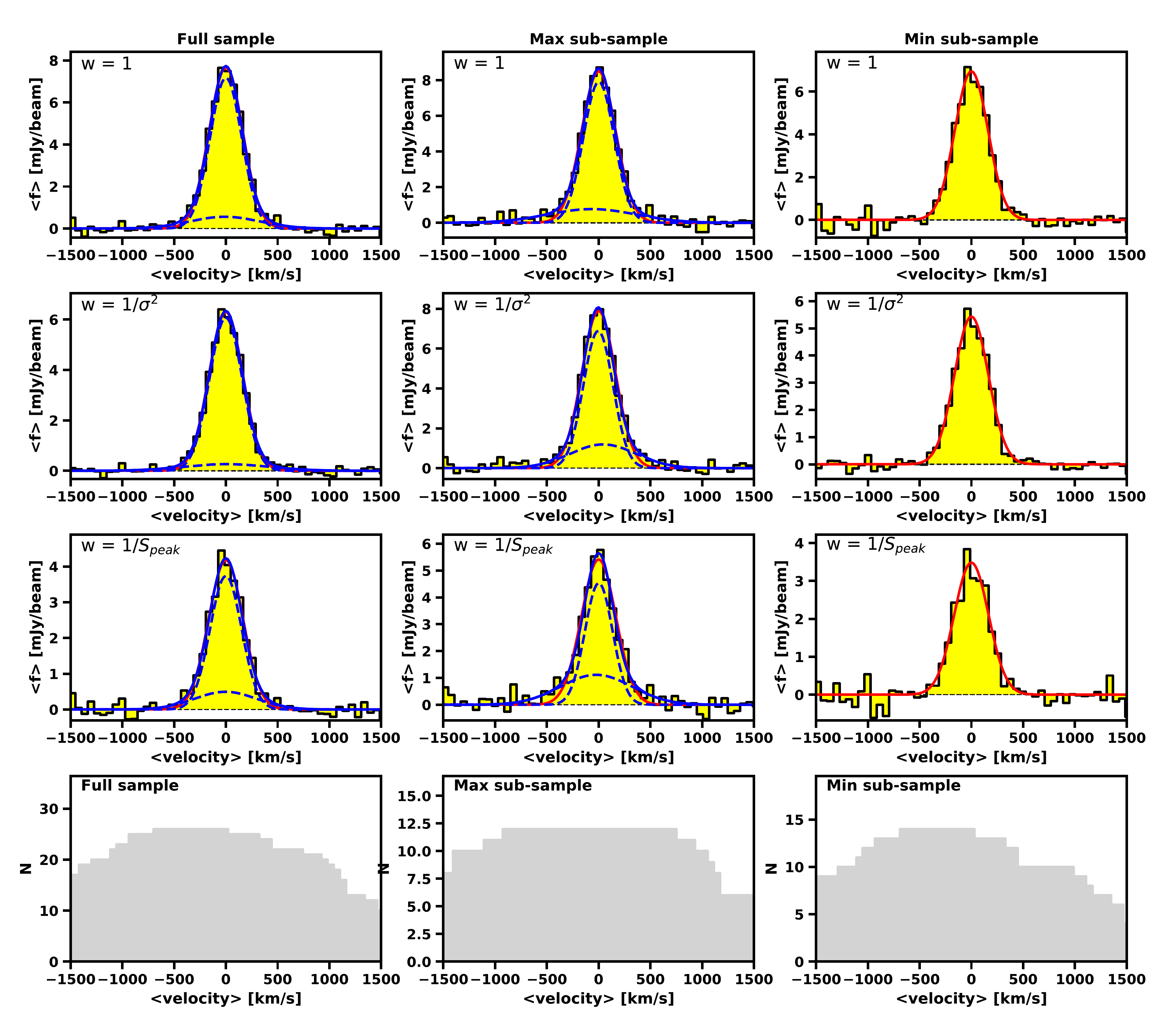}
      \caption{Results from our 1-D stacking analysis. {\em (left)} The stacked spectra of the full sample,{\em (middle)} the {\em max sub-sample}, and {\em (right)} the min-sub-sample, using weights of $w = 1$ (1st row), $w = 1/\sigma_{\rm rms}^2$ (2nd row), and $w = 1/S_{\rm peak}$ (3rd row). On the last row we give the number of sources per channel of the stacked line.}\label{fig:1Dstackresults}   	
    \end{center}   
\end{figure*}

In the first column of Fig.~\ref{fig:1Dstackresults} we show the 1-D stacked spectra for the full sample, for the three different weights used. At the bottom of the figure we also show the number of sources included in each channel. The results from fitting the stacked line are given in Table~\ref{tab:results_cii_1D}. Including a broad component in the fit improves the $\chi^2$ of the fit by 24, 23, and 14\% for stacks with weights of $w = 1$, $w = 1/\sigma_{\rm rms}^2$, and $w = 1/S_{\rm peak}$, respectively. However, the fitted parameters for the broad component are not well constrained (see Table~\ref{tab:results_cii_1D}). \\
\indent 
We estimate the significance of the excess emission not accounted for by the single component fit to the line. For this we subtract the single component fit from the spectrum, and take the sum of the residuals within $\pm800$\,km/s. We find a significance of 0.4--0.7$\sigma$, depending on the weighting scheme used (see Table~\ref{tab:sig}).

\begin{figure*}[t]
  \begin{center}
      \includegraphics[width=\textwidth]{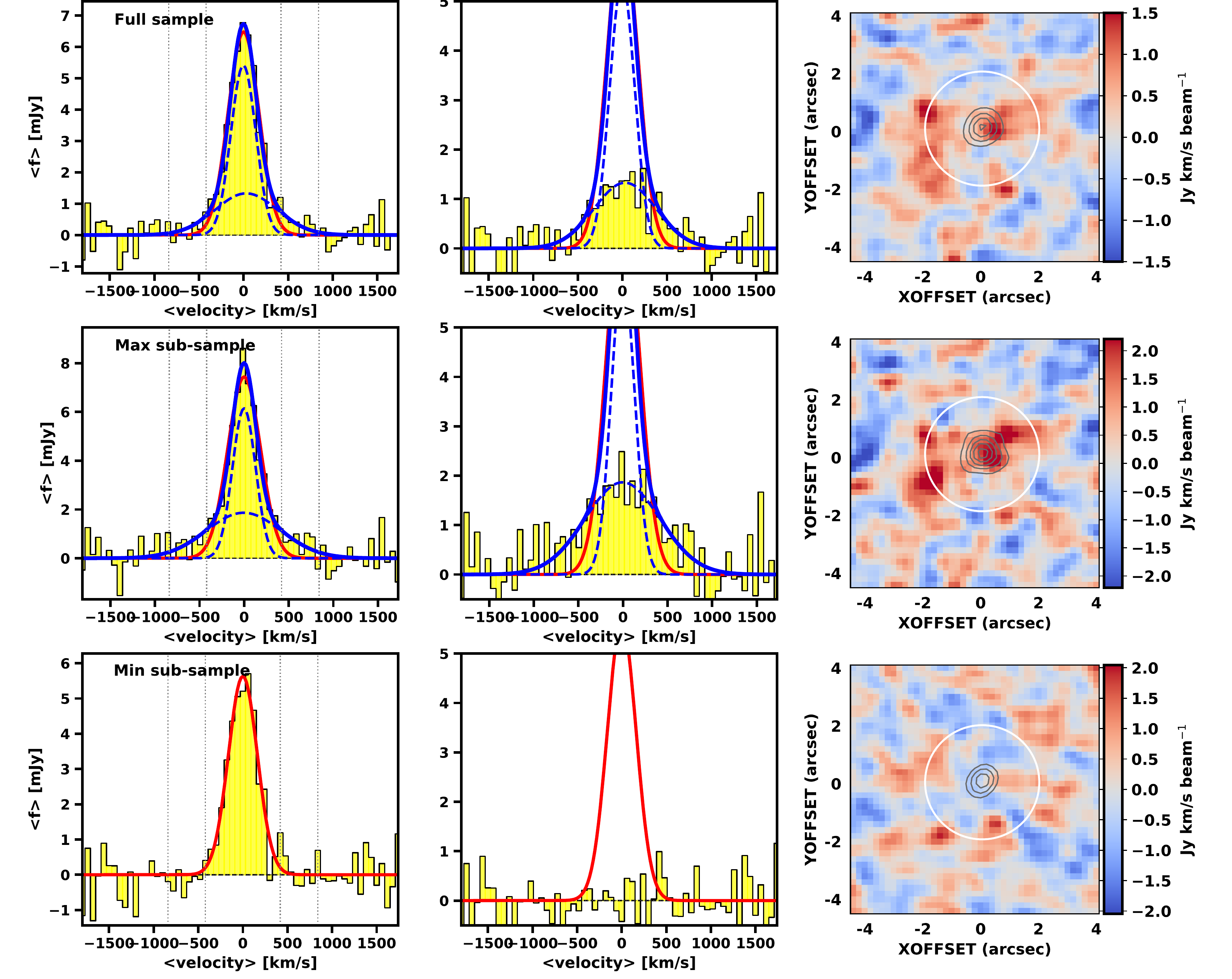}
      \caption{Results from our 3-D stacking analysis, for $w=1$. From left to right: the \cii line extracted
from the stacked cube within a 2\arcsec radius from the centre, a zoom-in view of the broad line component once the narrow component has been removed, combined image of the \cii line wings from $\pm$\,420--840\,km\,s$^{\rm -1}$ (see vertical dotted lines in the spectrum), for the full sample {\em (1st row)}, the {\em max sub-sample} {\em (2nd row)}, and {\em min sub-sample} {\em (3rd row)}. We also show the one (red curve) and two-component fits (blue solid and dashed curves) to the stacked line. The extent of the narrow component emission is shown with the black contours in the combined images, and the white circle corresponds to the area within a 2\arcsec radius from the centre.}\label{fig:stackresults}   	
    \end{center}   
\end{figure*}

In Fig.~\ref{fig:stackresults}\,{\em (top panel)} we show the stacked cubes of the full sample for the non-weighted stacks. Also shown are the one-component (red) and two-component (blue) fits to the line. The results of the two-component fits to the stacked line are given in Table~\ref{tab:results_cii_3D}. In the right of Fig.~\ref{fig:stackresults} we also show the collapsed image of the wing channels, covering velocities of $\pm 420-840$~km\,s$^{\rm -1}$. In this case, including a broad component improves the $\chi^2$ of the fit by 46\%, 15\%, 34\% for weights of $w = 1$, $w = 1/\sigma_{\rm rms}^2$, and $w = 1/S_{\rm peak}$, respectively. Following the methods described earlier, we calculate the significance of the excess emission to be 1.1--1.5$\sigma$, depending on the weighting scheme used (see Table~\ref{tab:sig}). \\
\indent
The 3-D stacked line differs to that from our 1-D stacking analysis. This is due to the fact that in our 3-D stacking analysis we use a larger integration area (2\arcsec) than that used in the 1-D stacking analysis, where only the area with positive signal was included. 
 Consequently, the 3-D stacked line, includes faint extended signal missed by the moment-0 2$\sigma$ regions used for the 1-D stacking analysis.
 Comparing the results of the 1-D stacked spectra with those of the 3-D stacked spectra it becomes clear that the broad component in the 1-D stacking is poorly constrained, due to the lack of the more extended signal that is included in the 3-D stacking.

\subsection{A sub-sample with outflow signatures}\label{subsect:res_sub-sample}

\begin{figure}
    \centering
    \includegraphics[width=0.8\columnwidth]{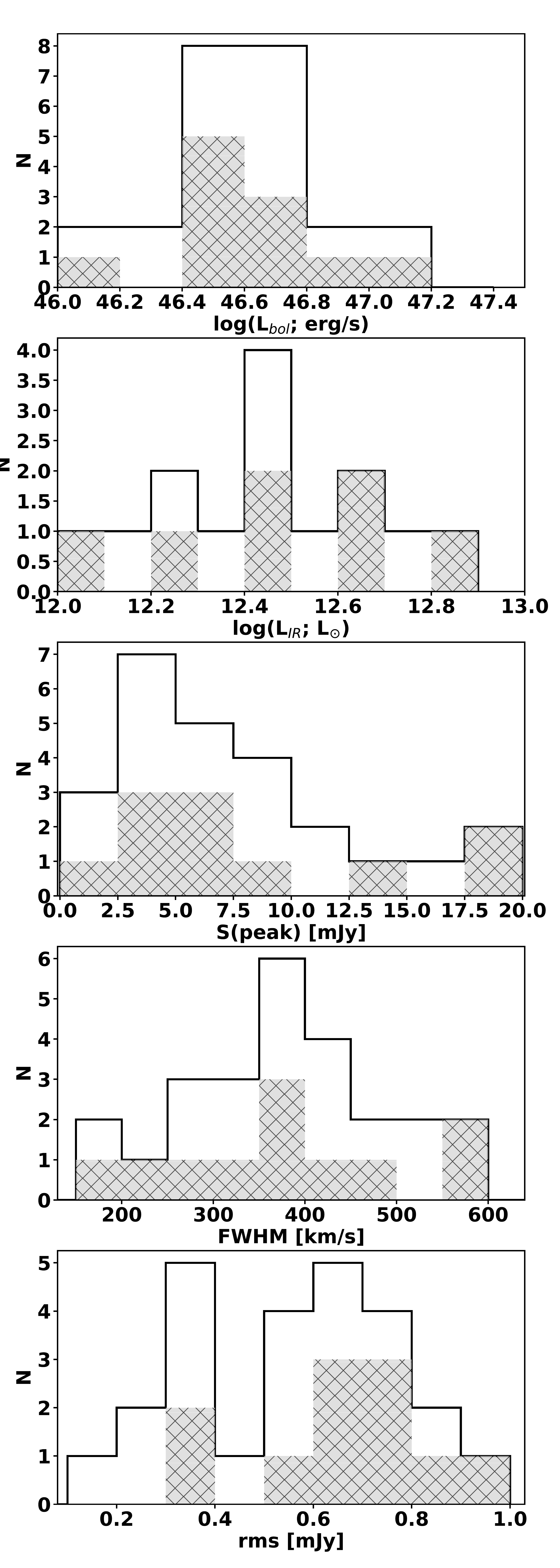}
    \caption{Here we show the distribution of the properties of the sources in the {\em max sub-sample} (filled in regions) in comparison to the full sample of this study (main histogram). From top to bottom we compare: the bolometric luminosity of the quasars (L$_{\rm bol}$), the IR (8--1000$\mu m$) luminosity of the galaxies (L$_{\rm IR}$), the $S_{\rm peak}$, the FWHM, and the rms of the of the individual \cii lines (before the velocity rebinning). It is evident that the {\em max sub-sample} covers a wide range of values in all properties, and is not skewed towards the brightest quasars, specific line-widths or line strength. The {\em max sub-sample} does seem to prefer higher rms values, but this is restricted by small number statistics.}
    \label{fig:histogram_max}
\end{figure}

In addition to simply stacking the full sample in search for an outflow signature, we have also developed a method for determining a sub-sample that maximises the possibility of detecting an outflow signature in the stacked spectrum. Under normal circumstances it is assumed that all stacked
sources have the same properties, but in the case of outflows, the orientation of a bi-polar
outflow will impact on how it might be detected. If the outflow is oriented in or close to the plane of the sky, the radial component of the velocity will be relatively small and likely remain undetected. This means that even if all sample sources would have a high-velocity outflow, we would
expect a fraction of these not to contribute to the stacked signal.\\
\indent
The sub-sample selection is done by randomly selecting sub-sets of $n$ sources, where $n$ is in the range 3$-$25, and repeated 10\,000 times.
For each of the 10\,000 randomly selected sub-samples, we perform the same 
stacking analysis as described above. From the resulting stacked spectrum we select 
the ``line-free'' channels, defined to be at a distance of 2$\,\times\,\rm FWHM$ of the 
center of the main line component. These are channels were we can expect to see only emission 
due to an outflow component. We then integrate over these channels at different radii from the center, within the range of 0.2--3\arcsec in steps of 0.2\arcsec.
The integrated flux density for each integration radius is saved as a grade for the 
sub-sample.\\ 
\indent
Once this is done for all 10\,000 randomly selected sub-samples, each source in our sample
is ranked based on the total sum of the grades of the sub-samples it was in at all 
integration radii. Sources with no outflow signal will be mostly in sub-samples 
with low grades, and will therefore have an overall low rank. 
We take the mean rank of all sources and define it as the lower limit for selection of the best 
sub-sample, i.e., the sub-sample of sources most likely to have signal from an outflow component ({\em max sub-sample}).
We also define the sub-sample of least likely sources to have signal from an outflow component ({\em min sub-sample}), picked 
to have a rank below or equal to the mean. \\
\indent
The {\em max sub-sample} consists of 12 sources highlighted with boldface in Table~\ref{tab:results_cii}.  
We note that the subsampling analysis was performed for both the 1-D and 3-D stacking
analysis and returned the same sub-sample for both cases. 
These sources cover
the same range in bolometric quasar luminosities, redshifts, IR luminosities and \cii line properties as the full sample (see Fig.~\ref{fig:histogram_max}). \\
\indent
The stacked line of the {\em max sub-sample} is shown in the second column of Fig.~\ref{fig:1Dstackresults} and the second row of Fig.~\ref{fig:stackresults}.
An outflow component is found in both the 1-D and 3-D stacking results, for all three weighting schemes. In the case of the 1-D stacking analysis we find that the fit when including a broad component, is improved by 48\%, 51\%, 46\% for weights of $w = 1$, $w = 1/\sigma_{\rm rms}^2$, and $w = 1/S_{\rm peak}$, respectively. We find that there is excess emission of 1.1--1.5$\sigma$, depending on the weighting scheme used (see Tables~\ref{tab:results_cii_1D}\,\&\,5). In the case of the 3-D stacking analysis the presence of a broad component is more pronounced. We find that including the broad component improves the fit by 62\%, 57\%, 48\% for weights of $w = 1$, $w = 1/\sigma_{\rm rms}^2$, and $w = 1/S_{\rm peak}$, respectively. Furthermore, the significance of excess emission rises to 1.2--2.5$\sigma$ (see Tables~\ref{tab:results_cii_3D}\,\&\,\ref{tab:sig}).
The collapsed image (over $\pm$420-840\,km\,s$^{-1}$) in Fig.~\ref{fig:stackresults}, shows spatially extended emission, larger than what is seen for the narrow component (see contours), reaching $\sim$2\arcsec (see Fig.~\ref{fig:stackresults}). \\
\indent
As discussed earlier, the 3-D stacking results are better at constraining the broad component signal compared to the 1-D stacking, due to the fact that it includes the more extended faint emission excluded in the 1-D stacking. Therefore, for the rest of this paper we only consider the 3-D stacking results.\\
\indent 
We estimate the mean outflow rate ($\dot{M}_{outf}$) for the {\em max sub-sample}, based on the $w=1$ 3-D stacking results. We use the equation of \cite{Hailey-Dunsheath10} to calculate the outflow gas mass from the integrated \cii luminosity of the broad component, which includes dependencies on the gas density ($n$), temperature (T), and abundance of C$^+$ ions (X$_{\rm C^+}$). Following \cite{Maiolino12}, and \cite{Cicone15}, we assume that X$_{\rm C^+} = 1.4\times10^{-4}$, T\,$=$\,200\,K, and $n>>n_{crit}$ ($n_{crit} \sim 3\times10^3 \rm cm^{-3}$), which are typical of photo-dissociated regions. We calculate an estimate of the velocity ($\upsilon_{outf}$) of the outflow by taking $\upsilon_{outf}=0.5\times$FWHM, and take the radius of the extent of the outflow emission ($R$). By assuming that the velocity of the outflow is constant across the outflow, the dynamical time of the outflow can be defined as $\tau_{dyn} = R/\upsilon_{outf}$. Therefore, $\dot{M}_{outf} = M_{outf}/\tau_{dyn} = M_{outf}\times\upsilon_{outf}/R$. For the mean redshift of 6.236, a radius of $R = 11.5kpc$ (2\arcsec), and $\upsilon_{outf} = 493$\,km\,s$^{-1}$ calculate that $\dot{M}_{outf} = 45\pm21$\,M$_\odot$\,yr$^{-1}$. Since the properties of the broad component do not change significantly for the different weights, the calculated mass outflow rate will be similar for all weights.\\
\indent
To determine if a specific source in our {\em max sub-sample} is driving the outflow component emission and/or the shape of the extended emission, we perform a test. We repeat our stacking analysis as many times as the number of sources 
in the {\em max sub-sample}, and each time remove one of the sources. As a result we have the stacked lines and collapsed images for each source's exclusion from the sub-sample. If one source is driving the observed results of the {\em max sub-sample}, then the stack without it will show significant differences to that of the {\em max sub-sample}.
We find that none of the sources cause a significant difference in the extent of the high-velocity emission seen in the collapsed images, and none are specifically driving the shape and strength of the outflow component of the stacked \cii line. Specifically, the FWHM, $S_{\rm peak}$ values of the broad component remain within 20\% of those found for the stack of the complete {\em max sub-sample}, and remain within the error of the fitted values (see Table~\ref{tab:results_cii_3D}).

\begin{table*}
	\begin{center}
	\caption{Results of our 3-D stacking analysis. Shown are the peak flux density ($S_{\rm peak}$), 
	FWHM, integrated flux density ($S_{\rm int}$), and central frequency ($\nu_{\rm cen}$) of both the narrow and broad component of the stacked lines.}\label{tab:results_cii_3D}
		\begin{tabular}{lcccccccccc}
			\hline \hline
		
             & \multicolumn{4}{c}{Narrow component} & \multicolumn{4}{c}{Broad component} \\
			 & $S_{\rm peak}$ & FWHM & $S_{\rm int}$ & $\nu_{\rm cen}$ & $S_{\rm peak}$ & FWHM & $S_{\rm int}$ & $\nu_{\rm cen}$\\
				&  [mJy]   &  [km\,s$^{-1}$] & [Jy km\,s$^{-1}$] & [GHz] &  [mJy]	& [km\,s$^{-1}$] & [Jy km\,s$^{-1}$] & [GHz] \\
			\hline
			\underline{Full sample} \\
             $w = 1$ &  $5.41\pm0.89$ & $328\pm41$ & $1.9\pm0.4$ & $-2\pm10$ & $1.33\pm0.92$ & $820\pm261$ & $1.1\pm0.9$ & $30\pm65$ \\ 
             $w = 1/\sigma_{\rm rms}^2$ &  $6.54\pm0.43$ & $373\pm23$ & $2.6\pm0.2$ & $2\pm6$ & $0.37\pm0.44$ & $1029\pm614$ & $0.5\pm0.6$ & $-100\pm213$ \\  
             $w = 1/S_{\rm peak}$ & $2.93\pm1.05$ & $300\pm82$ & $0.9\pm0.4$ & $-19\pm22$ & $1.62\pm1.08$ & $724\pm211$ & $1.2\pm0.9$ & $56\pm75$ \\ 
			
			\underline{\em max sub-sample} \\
			$w=1$ &  $6.14\pm0.69$ & $305\pm36$ & $2.0\pm0.3$ & $-1\pm10$ & $1.87\pm0.68$ & $986\pm210$ & $2.0\pm0.8$ & $-5\pm58$ \\
			$w = 1/\sigma_{\rm rms}^2$ & $7.53\pm0.4$ & $348\pm22$ & $2.8\pm0.2$ & $2\pm7$ & $0.81\pm0.33$ & $1427\pm408$ & $1.2\pm0.6$ & $-100\pm138$ \\
			$w = 1/S_{\rm peak}$ & $3.05\pm0.85$ & $182\pm62$ & $0.6\pm0.3$ & $-24\pm20$ & $3.31\pm0.69$ & $775\pm116$ & $2.7\pm0.7$ & $10\pm38$ \\ 
						
			\underline{\em min sub-sample} \\
			$w=1$ & $5.63\pm0.31$ & $383\pm24$ & $2.3\pm0.2$ & $0\pm10$ &  .. & .. & .. & ..\\ 
         	$w = 1/\sigma_{\rm rms}^2$  & $6.43\pm0.21$ & $390\pm14$ & $2.7\pm0.1$ & $0\pm6$   & .. & .. & .. & .. \\ 
            	$w = 1/S_{\rm peak}$ &  $3.61\pm0.39$ & $378\pm47$ & $1.4\pm0.2$ & $0\pm20$ & .. & .. & .. & .. \\
			\hline
		\end{tabular}
	\end{center}
\end{table*}

\begin{table}
	\begin{center}
	\caption{The significance of the excess emission over the one component fit to the spectra (red curve in Fig.~\ref{fig:1Dstackresults}~\&~\ref{fig:stackresults}), for the Full sample and {\em max sub-sample} in all three weighting schemes used.}\label{tab:sig}
		\begin{tabular}{lcccc}
			\hline \hline
		
			 &  1-D stacking & 3-D stacking \\
			\hline
            \underline{Full sample} \\ 
            $w = 1$ & 0.7$\sigma$  & 1.5$\sigma$ \\
            $w = 1/\sigma_{\rm rms}^2$  & 0.6$\sigma$ & 1.3$\sigma$  \\
            	$w = 1/S_{\rm peak}$ & 0.4$\sigma$ & 1.1$\sigma$ \\ 
			\hline  
            \underline{{\em max sub-sample}} \\ 
            $w=1$ & 1.3$\sigma$ & 2.1$\sigma$ \\ 
            $w = 1/\sigma_{\rm rms}^2$ & 1.5$\sigma$ & 2.5$\sigma$  \\ 
            $w = 1/S_{\rm peak}$ & 1.2$\sigma$ & 1.2$\sigma$ \\	
			\hline
		\end{tabular}
	\end{center}
\end{table}

If we extract the line only from the central pixel then 
there is no evidence for an outflow, in agreement 
with the results by \cite{Decarli18}. Taking the central pixel is equivalent to taking the flux within the beam for point like sources. This suggests that the emission of the outflow is spatially extended beyond the beam, as indicated by the collapsed image of the line wings in Fig.~\ref{fig:stackresults}.\\
 \indent
We have also looked into the stack of the sources 
excluded from the {\em max sub-sample} described above (right column of Fig.~\ref{fig:1Dstackresults} and lower row of 
Fig.~\ref{fig:stackresults}). A stack
for these sources shows no evidence of an outflow 
component, and the line wing images show no significant 
traces of emission. This result also adds to the reasonableness of 
our {\em max sub-sample} selection, since it does not show any outflow signal, but also 
the  average signal is not negative which could be the case if the signal in the 
{\em max sub-sample} was artificial.

In this analysis we have assumed that the outflow emission signature has the spectral profile of a 
single broad Gaussian, which might not be the case for some or all of the quasars in our sample.
Previous individual detections of outflows reported in the literature using e.g., CO, HCN, or \cii, 
show that the majority of the lines have nearly symmetric, broad emission lines, with a subset of results showing outflows 
dominated by emission only on the red or blue side of the main component \citep[e.g.,][]{Cicone14,Janssen16}.  
Apparent individual emission line components with a high velocity offset from the dominant line component could 
also be interpreted as an outflow signature \citep[e.g.,][]{Fan18}. Furthermore, we have assumed that the broad component found 
is originating from the high-velocity gas of an outflow. However, it is possible that the high-velocity emission is not due to 
an outflow but for example, a gas flow between two interacting galaxies \citep[for a detailed discussion see][]{Gallerani18}. 
We know of one source included in our sample (PJ308-21) that has complex high-velocity structure \citep[e.g.][]{Decarli17}. 
However, when this source is removed from our stacking analysis, the observed broad component and the spatial extent of the wings do not change significantly.

\subsection{Testing the feasibility of detecting an outflow signal with our method.}

To determine if the methods of our analysis are indeed capable of detecting an outflow component, we test their performance on mock samples with and without an outflow component. To test if our method can successfully retrieve the broad component signal, we perform our 1-D stacking analysis on a mock sample with \cii lines that include a weak broad component. Based on the extracted spectra of our the observed sample, we use the one-component line fits of the observed lines, to which we add a broad line component of similar properties to what we find in our stacking results, with 2 times the width and 0.1 times the peak flux density of the main line component. For 10\,000 iterations, we add random noise to the individual \cii lines, based on the range of the observed rms, and stack following our 1-D stacking analysis. We then determine if we can detect excess emission due to a broad component in the stacked spectra, following the method described in section~\ref{subsec:res_fullsample}. We find excess emission due to a broad component with only a weak significance of $<3\sigma$, with 75\% of iterations having an excess emission of $1-2\sigma$ for stacks with weights of $w = 1$ and $w = 1/\sigma^2_{rms}$. However, a weight of $w = 1/S_{peak}$ seems to suppress the signal, with the detectability of excess emission dropping to 20\% for a $1-2\sigma$ significance (see figures in Appendix~\ref{appendix}).
	It is worth noting however, that if assuming a stronger initial broad component, with 3 times the width and 0.2 times the peak flux density of the main line component, then our method retrieves a strong broad component in the stacks. In this case an excess emission of $>5\sigma$  for $>99\%$ of iterations, for stacks with weights of $w = 1$ and $w = 1/\sigma^2_{rms}$, and $>3\sigma$ for 35\% of iterations for $w = 1/S_{peak}$. \\
\indent
To test if it is possible that the outflow component found in our results could be created by stacked noise and is not a real signal, we perform a similar simulation. Based on the one-component fit to the \cii lines of our sample, we create a mock sample with the same line properties, and without an added outflow component. We then add random noise to each based on the range of the observed rms. We perform our 1-D stacking analysis on this mock sample 10\,000 times, each time re-applying randomly selected noise levels. Then we attempt a two-component fit to each stacked line and determine if a broad line component is found and if it has comparable properties to the one found in our analysis of the observed data. This is done for all three weighting schemes. 
We find that, depending on the weight scheme used, 14-17\% of the iterations return a $>0.4\sigma$ excess signal in the stacked lines (see Appendix~\ref{appendix}).

\section{Discussion}
\label{sec:disc}
\subsection{Spectral line stacking methods}
\label{sec:method_impl}

At the high redshifts covered in our study the spatial information available is limited. Therefore, 
our analysis will primarily be sensitive to the information that can be derived from the velocity characteristics of the observed lines. Hence our type of analysis is relevant for outflows that have velocities 
higher than that of the host galaxy emission, and low-velocity outflows will not be detected.  \\
\indent
As the aim of the stacking is to search for a faint broad emission line component, it is imperative to understand the effect of stacking data where the main (bright) component has varying line widths.  
If spectral lines, such as those of the data used here, are simply stacked, it could result in the detection of an artificial broad component. 
To test the effect of the varying FWHM values of the lines stacked on the final result, we create a mock sample with the observed line properties of the observed sample, without adding an outflow component. We perform our stacking analysis on this mock sample 10\,000 times, each time re-applying randomly selected noise levels within the range of observed rms. For each iteration we perform both 
straightforward stacking, and stacking using our rebinning method. For the resulting stacked spectra we attempt two-component fitting to determine if a significant broad component is found. We find that for the simple/straightforward stacking analysis, excess emission due to a broad component is found at $>2\sigma$ 2-5\% of the time. However, we find that these cases are due to the noise of the stacked spectrum. If we repeat the exercise without including noise in the spectra, then a broad component is not found. It is worth noting that even if a significant broad component is not fitted in the majority of the cases, there is still low significance ($<1\sigma$) excess emission in the line wings that the single component gaussian cannot fit (see examples in Appendix~A). \\
\indent
Recently, there have been two other studies looking at similar samples of high-$z$ quasars with \cii \citep[][]{Decarli18,Bischetti18}. In \cite{Decarli18} a sample of $z\sim6$ quasars observed with \cii is presented, and a brief analysis on the presence of outflows is included. 
The overall analysis is not that different to our own, as they also use velocity rebinning before stacking, but have only stacked the spectra extracted from the central pixel of each source. They report no evidence for a broad component in the stacked spectra, which is consistent with the results of our 1-D stacking analysis of the Full sample. However, as we have demonstrated in this paper, it is imperative to take into account the emission from the entire extent of the galaxy, as the outflow emission may be originating from an extended region. \\
\indent
\cite{Bischetti18} used archival data of quasars in the wider redshift range of $4.5<z<7.1$, to search for outflow signatures in the stacked spectra. However, it is not clear how the archive data have been treated; as noted previously, we needed to manually re-calibrate 70\% of the archive observations of our sample.
The individual spectra were integrated over a region equivalent of four times the beam size, and were stacked without velocity rebinning but with a weighting scheme of $1/\sigma_{\rm rms}^2$ for each channel of each spectrum. 
Evidence for a broad outflow component are found, in agreement with our conclusions, but with a stronger significance. The higher significance of the broad component found in the \cite{Bischetti18} analysis could be due to the absence of velocity rebinning to account for the range in the FWHM values of the lines stacked, which could result to an overestimate of the flux and FWHM of the broad component. However, it could also be due to the larger sample used in their analysis.\\
   
\indent
Even though velocity re-binning is important in order to not introduce artificial signal to the broad line emission, it also implies that the line width of the broad line component is roughly proportional to that of the main component. Consequently, our analysis can detect a broad component, but it will not be based on the absolute value of the line width, but rather on the mean of the velocity axis of the spectra stacked. However, current available spectroscopic data lack the sensitivity to fully detect individual outflows, and spectral stacking is the only available method at present to determine the presence of outflow signatures.

\subsection{Should we expect to see outflows in all high-$z$ quasars?}

As discussed previously, we find only a tentative signal for an outflow component when stacking 
the full sample of $z\sim6$ quasars, but find that for a specific subsample of sources we can detect a 
significant outflow component. This would suggest that either not all $z\sim6$ quasars have 
outflows (at least at the time of observation), or there are effects that inhibit 
the observability of present outflows. \\
\indent
Recently, a study by \cite{Barai18} using zoom-in hydrodynamical
simulations to determine the effect of outflows on the host galaxy and
surrounding environment of quasars at $z\geq6$, has demonstrated that
outflows could indeed be a dominant feature of high redshift quasars. Even
though the quasar host galaxies at $z\sim6$ are accreting significant amounts
of cosmic gas, AGN feedback succeeds in reducing the inflow by $\sim12\%$,
with $\sim20\%$ of the quasar outflows having speeds greater than the escape
velocity of 500\,km\,s$^{\rm -1}$, subsequently succeeding in ejecting gas out of the host
galaxy, and regulating the on-going star formation \citep{Barai18}. \\

\subsubsection{Outflow orientation}\label{sec:orientation}
The presence, strength, and width of a broad line component due to outflow emission will be dependent on the orientation of the outflow with respect to the observer's line of sight. Indeed, in a recent study by \cite{RobertsBorsani18} investigating 
cold gas inflows and outflows around galaxies in the local Universe, it was found that outflows 
were detected only for outflow angles $<$\,50\degr 
in respect to the line of sight, while for $>$\,60\degr it was {\bf only} possible to detect inflows.\\
\indent
To test how orientation angles could affect the observability of an outflow
component in the stacked spectra we run a simple Monte-Carlo test. 
In its most simplistic form, we assume that all galaxies in the sample 
have the same values of peak flux density and FWHM of their lines and have
the same outflow component. \\
\indent
We first define a line profile that is a combination of a narrow and broad component. 
For the narrow component we assume $S_{\rm peak}=10$\,mJy and FWHM$ = $400\,km\,s$^{\rm -1}$. For the broad component, 
we assume five different FWHM values (800, 900, 1000, 1100, 1200\,km\,s$^{\rm -1}$) and repeat the test for each.
 \\
\indent
For each of the broad component FWHM values, we run 1000 iterations, in each of which we select 30 random angles between 10--80\degr (simplistic assumption to avoid a blazar, 0\degr, and obscured quasar, 90\degr, types).
We combine the selected random orientation angles with the defined line profile to create a mock sample of 30 galaxies, and add random noise with rms of $\sim$0.5\,mJy, similar to what is seen in the observed sample. We then stack the spectra following the same methods as described in sect.~\ref{sect:stackingmethods}. We fit the resulting stacked spectral line with a single and double-component Gaussian function, and calculate the respective $\chi^2$ values of the fits, and the significance of the broad component emission. \\
\indent
We find that an excess emission due to a broad component, of  $>2\sigma$ significance is retrieved for $27-50\%$ of the iterations, for the cases where we have assumed an initial broad component with FWHM$>$1000\,km\,s$^{\rm -1}$. For an initial broad component with FWHM of 800 and 900\,km\,s$^{\rm -1}$, a broad component in the stack with $>1\sigma$ is found in $>60\%$ of iterations (see Figures~A.4~\&~A.6 in Appendix A).

\subsubsection{Choice of outflow tracers} \label{sec:dif_tracers}
The choice of a tracer is an important part in all studies aiming to determine the presence and 
properties of outflows in galaxies. The recent studies on the presence of outflows in the most distant quasars (this work; \citealt{Decarli18}; \citealt{Bischetti18}) all show different results, and none 
can reproduce the very strong signal seen in SDSS J1148+5251 \citep[e.g.][]{Cicone15}. Therefore, it is important to consider if \cii is indeed a suitable tracer of outflows.
Are the weak outflow signatures that we observe due to the choice of \cii as a tracer? Is there 
a better tracer for outflows?

\cii 158$\mu$m is a tracer of
atomic gas at typical densities of $2.8\times10^3$\,cm$^{-3}$ with temperatures of $\sim$\,92\,K. It conveniently falls within the ALMA bands at $z\sim6$, and therefore is easy to observe and proves a very useful tracer, especially for quasar studies (see review by \citealt{CarilliWalter13}). However, at lower redshifts, there is a variety of more commonly used gas tracers used to
detect outflow signatures of AGN, such as $^{12}$CO tracing molecular gas.

\cii emission can originate from ionised, atomic, and molecular regions; however, 
both low and high redshift observational and theoretical work argue that the majority of 
\cii emission (70--90\%) originates from neutral atomic gas \citep[e.g.,][]{Croxall17,Olsen15, 
Lagache18}. Although, there has been observational evidence of \cii emission from ionised 
regions \citep[e.g.,][]{Contursi13, Decarli14}. \cii emission has also been found to trace molecular gas that is ``CO-faint'' due to the disassociation of the CO molecules from far-unltraviolet emission \citep[e.g.,][]{Jameson18}. 
However, \cite{Lagache18} highlight that the \cii emission at redshifts $z>6$ can be strongly affected 
by attenuation from the cosmic microwave background; although it is mostly warm and low density gas that is affected by this.  
\cii 158\,$\mu$m has been successful in detecting outflows in a few nearby galaxies
\citep[e.g.,][]{Contursi13,Kreckel14}, but at redshifts of 
$z\sim6$ the only example of a quasar outflow detected with \cii is J1148+5251 
\citep[e.g.,][]{Maiolino12,Cicone15}. 

It is possible that CO cannot survive in the
outflow regions; however, CO(5-4) has been seen in an ionised outflow 
region \citep[e.g.,][similar result found in Fogasy et al., in prep]{Brusa18} at high redshifts, and there are plenty of nearby galaxies with resolved molecular outflows traced with multiple CO transitions \citep[e.g.,][]{Alatalo11, Aalto12, Cicone12, Cicone14, Feruglio15, Veilleux13, Veilleux17}. 

Although H{\sc $\alpha$} and [O{\sc iii}], as well as 
X-ray emission can directly trace the shocked and ionised regions caused by the outflows, and hence may be a more direct probe, such observations are currently impossible at such high redshifts. There is currently only one tentative detection of hot gas in the X-ray \citep[SDSS J1030+0524,][]{Nanni18}. However, future JWST projects may be able to probe the ionised gas in high redshift outflows.

Overall, due to the lack of systematic studies on the topic it remains unclear if \cii - so far the most commonly studied outflow tracer at high-$z$ - is a good tracer of outflows in distant galaxies. However, with the sensitivity of ALMA 
it is possible to have systematic studies in the future, to address this.

\subsection{The presence of outflows in high-redshift quasars.}\label{subsec:disc_comp}

Individual studies of high-redshift quasars have revealed some evidence for
outflows based on \cii and/or CO observations. At a redshift of $z\sim6.4$,
the luminous quasar SDSS J1148+5251 shows a strong \cii broad component
indicating the presence of an outflow with velocities of up to 1400\,km\,s$^{\rm -1}$ and
extending to distances of $\sim30$\,kpc \citep[]{Maiolino12,Cicone15}. At the
lower redshifts of $z\sim1.4$, $z\sim1.6$, and $z\sim3.9$, \cite{Vayner17}, \cite{Brusa18}, and \cite{Feruglio17} 
report the detection of a molecular outflow traced with CO. In the first two cases the outflow velocities are 
600--700\,km\,s$^{\rm -1}$, while in the third
case the outflow velocity reaches up to $\sim$1300\,km\,s$^{\rm -1}$. At redshifts of $z\sim1-3$ there are also numerous 
detections of ionised outflows traced with [O{\sc iii}] \citep[e.g.,][]{Harrison12,Harrison16,Brusa16,Carniani16,Vayner17}.

In studies of lower redshift AGN outflows of $z<0.2$ using [O{\sc iii}] ionised gas tracer, there is 
a prevalence of outflows in both luminous AGN \citep[e.g.,][]{harrison14} and among 
ULIRGs hosting AGN \citep[e.g.,][]{Rose18, Tadhunter18, Spence18}, 
with typical line widths of $\simeq600-1500$\,km\,s$^{\rm -1}$ and $\simeq500-2500$\,km\,s$^{\rm -1}$. 
Similarly, molecular outflow studies of nearby quasar galaxies using CO, HCN, and other molecular tracers, 
have found outflows with velocities reaching $\simeq750-1100$\,km\,s$^{\rm -1}$ \citep[e.g.,][]{Cicone14,Aalto15,Veilleux17}.

For the max sub-sample we find a mean outflow velocity of $\sim$337--713\,km\,s$^{\rm -1}$, 
consistent with observations of both ionised and molecular outflows in single objects up to $z\sim2.5$.
Our results are also consistent with the range of mean outflow velocities reported in \cite{Bischetti18}.
We find that the ratio of the peak flux densities of the outflow over the main component 
found for our full sample and max-subsample at 2\arcsec\,
both range between $\simeq$\,0.1--0.5, and is highly dependent on the weighting scheme using. This is in agreement with the relative ratio of 
0.14 seen in J1148+5251 at $z\sim6$ \cite[e.g.,][]{Maiolino05,Cicone15}, while 
larger than seen in Mrk231 ($\simeq$\,0.067) when using the different tracer CO(1-0) in the nearby Mrk231 \citep[e.g.,][]{Cicone12}. 

In order to improve our understanding on the prevalence of outflows at $z\sim6$ and above we will need 
deeper observations of quasar samples in both \cii and CO, as well as tracers of the warm ionised medium. 
However, a multi-tracer approach is costly in terms of observing time, and thus the methods presented in this work provide a useful step towards designing the relevant observations.  
Large emission line surveys of high-redshift quasars can be combined with the sub-sampling method presented in Sect.~\ref{subsect:res_sub-sample}, to determine the more likely candidates to have outflows for deeper multi-phase observations. 
Apart from the ability to select a sub-sample of most likely sources to have an outflow, sub-sampling can also help to take into account potential effects such as orientation or selection effects.

%
%





\section{Conclusions}
 \label{sec:summary}
We have carried out a study of spectral stacking \cii
 ALMA archive observations of  $z\sim6$ optically selected
quasars, to search for faint outflow signal. We performed this analysis for both extracted spectra 
and full spectral cubes, allowing for the extraction of stacked spectra over different regions. We also performed a sub-sampling technique to identify the sub-sample of our sources that maximise the broad component signal when stacked.\\

\noindent Specifically:

\begin{itemize}
\item[$\bullet$]
We find tentative broad component signal for our full sample of 26 quasars. The stacked spectrum has a broad component of FWHM\,$>700$\,km\,s$^{-1}$ with a total line flux density of $S_{\rm int} > 0.5$\,Jy\,km\,s$^{-1}$. The excess emission, above a single component fit to the stacked line, corresponds to  0.4--1.5$\sigma$, depending on the weighting scheme used in the stack.
\item[$\bullet$]
We identify a sub-sample of 12 sources, which maximise the broad component signal when stacked. The
broad component detected in this sub-sample is characterised by an average FWHM\,$>775$\,km\,s$^{-1}$, and $S_{\rm int} > 1.2$\,Jy\,km\,s$^{-1}$ . The excess emission is detected at 1.2--2.5$\sigma$, depending on the weighting scheme used.
These results are consistent with results of individual sources at low and high redshifts. 
\item[$\bullet$]
We test if our methods could introduce artificial outflow signal when stacking, 
by creating mock sampled based on the observed \cii line properties of our sample. In this mock sample there is no broad component included. We find that the probability to 
falsely detect a broad component with an excess emission of $>0.4\sigma$ is $<$17\%.

\end{itemize}
Using spectral line stacking as a tool for this type of studies has the
following implications: 
\begin{itemize}
\item[$\bullet$]
Spectral line stacking is a meaningful tool for searching for outflows in the
spectra of quasars. Given that the redshift is well known because of the
bright line emission from the host galaxies, there will be no significant
velocity offsets that would dilute the emission from the fainter (broad)
component. However, it is important to take into account the varying line
width of the main line component in order to avoid introducing additional
broad spectral features. 
\item[$\bullet$]
As outflows are thought to be anisotropic emission that is typically seen as bi-polar, and for which the orientation is most likely random, it is expected that
at least a sub-set of the sample will not contribute significant or any emission from the outflow to the stacked spectrum. For instance, if the outflow is
aligned close to or on the line of sight, the projected radial velocity
might be the same or less than that of the main component.  
For a sample
typically of the size studied here, if the orientation is random, it would be
possible to obtain an outflow signal if the outflow velocity is larger than
twice the main component line width. To take into account the
fact that not all sources will be contributing significantly, we developed a
tool for identifying the sub-sample most likely to have outflow signatures in its stacked spectrum. 
\item[$\bullet$]
We have used the assumption that the outflow is characterised by a high
velocity component with a line width significantly larger than that of main
line, and such spectral signatures are seen in some local AGNs that have
detected outflows. However, it is possible that outflows may have other spectral characteristics, that will not be traced with our methods.
\end{itemize}
This type of study represents a step towards better insights in outflows and
possible feedback in massive galaxies, however, to gain full insights the
next steps will need to include direct detection of individual sources as
well as the detection of multiple outflow tracers.

\begin{acknowledgements}
We thank the anonymous referee for their instructive comments for the improvement of this paper. FS 
acknowledges support from the Nordic ALMA Regional Centre (ARC) node based at Onsala Space Observatory. 
The Nordic ARC node is funded through Swedish Research Council grant No 2017-00648. 
KK acknowledges support from the Swedish Research Council and the Knut and Alice Wallenberg Foundation. 
This paper makes use of the following ALMA data: ADS/JAO.ALMA\#2015.0.00606.S, ADS/JAO.ALMA\#2015.0.00997.S 
and ADS/JAO.ALMA\#2015.0.01115.S. ALMA is a partnership of ESO (representing its member states), 
NSF (USA) and NINS (Japan), together with NRC (Canada), MOST and ASIAA (Taiwan), and KASI (Republic of Korea), 
in cooperation with the Republic of Chile. The Joint ALMA Observatory is operated by ESO, AUI/NRAO and NAOJ.\\ 
This research made use of APLpy, an open-source plotting package for Python \citep[][]{Robitaille12}.
\end{acknowledgements}


\bibliographystyle{aa}
\bibliography{QSO_z6_outflow_stack}

\newcommand{\noop}[1]{}
\begin{thebibliography}{81}
\expandafter\ifx\csname natexlab\endcsname\relax\def\natexlab#1{#1}\fi

\bibitem[{{Aalto} {et~al.}(2012){Aalto}, {Garcia-Burillo}, {Muller}, {Winters},
  {van der Werf}, {Henkel}, {Costagliola}, \& {Neri}}]{Aalto12}
{Aalto}, S., {Garcia-Burillo}, S., {Muller}, S., {et~al.} 2012, \aap, 537, A44

\bibitem[{{Aalto} {et~al.}(2015){Aalto}, {Mart{\'{\i}}n}, {Costagliola},
  {Gonz{\'a}lez-Alfonso}, {Muller}, {Sakamoto}, {Fuller},
  {Garc{\'{\i}}a-Burillo}, {van der Werf}, {Neri}, {Spaans}, {Combes}, {Viti},
  {M{\"u}hle}, {Armus}, {Evans}, {Sturm}, {Cernicharo}, {Henkel}, \&
  {Greve}}]{Aalto15}
{Aalto}, S., {Mart{\'{\i}}n}, S., {Costagliola}, F., {et~al.} 2015, \aap, 584,
  A42

\bibitem[{{Alatalo} {et~al.}(2011){Alatalo}, {Blitz}, {Young}, {Davis},
  {Bureau}, {Lopez}, {Cappellari}, {Scott}, {Shapiro}, {Crocker},
  {Mart{\'{\i}}n}, {Bois}, {Bournaud}, {Davies}, {de Zeeuw}, {Duc}, {Emsellem},
  {Falc{\'o}n-Barroso}, {Khochfar}, {Krajnovi{\'c}}, {Kuntschner}, {Lablanche},
  {McDermid}, {Morganti}, {Naab}, {Oosterloo}, {Sarzi}, {Serra}, \&
  {Weijmans}}]{Alatalo11}
{Alatalo}, K., {Blitz}, L., {Young}, L.~M., {et~al.} 2011, \apj, 735, 88

\bibitem[{{Ba{\~n}ados} {et~al.}(2016){Ba{\~n}ados}, {Venemans}, {Decarli},
  {Farina}, {Mazzucchelli}, {Walter}, {Fan}, {Stern}, {Schlafly}, {Chambers},
  {Rix}, {Jiang}, {McGreer}, {Simcoe}, {Wang}, {Yang}, {Morganson}, {De Rosa},
  {Greiner}, {Balokovi{\'c}}, {Burgett}, {Cooper}, {Draper}, {Flewelling},
  {Hodapp}, {Jun}, {Kaiser}, {Kudritzki}, {Magnier}, {Metcalfe}, {Miller},
  {Schindler}, {Tonry}, {Wainscoat}, {Waters}, \& {Yang}}]{Banados16}
{Ba{\~n}ados}, E., {Venemans}, B.~P., {Decarli}, R., {et~al.} 2016, \apjs, 227,
  11

\bibitem[{{Ba{\~n}ados} {et~al.}(2014){Ba{\~n}ados}, {Venemans}, {Morganson},
  {Decarli}, {Walter}, {Chambers}, {Rix}, {Farina}, {Fan}, {Jiang}, {McGreer},
  {De Rosa}, {Simcoe}, {Wei{\ss}}, {Price}, {Morgan}, {Burgett}, {Greiner},
  {Kaiser}, {Kudritzki}, {Magnier}, {Metcalfe}, {Stubbs}, {Sweeney}, {Tonry},
  {Wainscoat}, \& {Waters}}]{Banados14}
{Ba{\~n}ados}, E., {Venemans}, B.~P., {Morganson}, E., {et~al.} 2014, \aj, 148,
  14

\bibitem[{{Barai} {et~al.}(2018){Barai}, {Gallerani}, {Pallottini}, {Ferrara},
  {Marconi}, {Cicone}, {Maiolino}, \& {Carniani}}]{Barai18}
{Barai}, P., {Gallerani}, S., {Pallottini}, A., {et~al.} 2018, \mnras, 473,
  4003

\bibitem[{{Barcos-Mu{\~n}oz} {et~al.}(2018){Barcos-Mu{\~n}oz}, {Aalto},
  {Thompson}, {Sakamoto}, {Mart{\'{\i}}n}, {Leroy}, {Privon}, {Evans}, \&
  {Kepley}}]{barcosmunoz18}
{Barcos-Mu{\~n}oz}, L., {Aalto}, S., {Thompson}, T.~A., {et~al.} 2018, \apjl,
  853, L28

\bibitem[{{Bischetti} {et~al.}(2019){Bischetti}, {Maiolino}, {Fiore},
  {Piconcelli}, \& {Fluetsch}}]{Bischetti18}
{Bischetti}, M., {Maiolino}, R., {Fiore}, S. C.~F., {Piconcelli}, E., \&
  {Fluetsch}, A. 2019, ArXiv e-prints, arXiv:1806.00786v4

\bibitem[{{Brusa} {et~al.}(2018){Brusa}, {Cresci}, {Daddi}, {Paladino},
  {Perna}, {Bongiorno}, {Lusso}, {Sargent}, {Casasola}, {Feruglio},
  {Fraternali}, {Georgiev}, {Mainieri}, {Carniani}, {Comastri}, {Duras},
  {Fiore}, {Mannucci}, {Marconi}, {Piconcelli}, {Zamorani}, {Gilli}, {La
  Franca}, {Lanzuisi}, {Lutz}, {Santini}, {Scoville}, {Vignali}, {Vito},
  {Rabien}, {Busoni}, \& {Bonaglia}}]{Brusa18}
{Brusa}, M., {Cresci}, G., {Daddi}, E., {et~al.} 2018, \aap, 612, A29

\bibitem[{{Brusa} {et~al.}(2016){Brusa}, {Perna}, {Cresci}, {Schramm},
  {Delvecchio}, {Lanzuisi}, {Mainieri}, {Mignoli}, {Zamorani}, {Berta},
  {Bongiorno}, {Comastri}, {Fiore}, {Kakkad}, {Marconi}, {Rosario}, {Contini},
  \& {Lamareille}}]{Brusa16}
{Brusa}, M., {Perna}, M., {Cresci}, G., {et~al.} 2016, \aap, 588, A58

\bibitem[{{Carilli} \& {Walter}(2013)}]{CarilliWalter13}
{Carilli}, C.~L. \& {Walter}, F. 2013, \araa, 51, 105

\bibitem[{{Carnall} {et~al.}(2015){Carnall}, {Shanks}, {Chehade}, {Fumagalli},
  {Rauch}, {Irwin}, {Gonzalez-Solares}, {Findlay}, \& {Metcalfe}}]{Carnall15}
{Carnall}, A.~C., {Shanks}, T., {Chehade}, B., {et~al.} 2015, \mnras, 451, L16

\bibitem[{{Carniani} {et~al.}(2016){Carniani}, {Marconi}, {Maiolino},
  {Balmaverde}, {Brusa}, {Cano-D{\'{\i}}az}, {Cicone}, {Comastri}, {Cresci},
  {Fiore}, {Feruglio}, {La Franca}, {Mainieri}, {Mannucci}, {Nagao}, {Netzer},
  {Piconcelli}, {Risaliti}, {Schneider}, \& {Shemmer}}]{Carniani16}
{Carniani}, S., {Marconi}, A., {Maiolino}, R., {et~al.} 2016, \aap, 591, A28

\bibitem[{{Cicone} {et~al.}(2012){Cicone}, {Feruglio}, {Maiolino}, {Fiore},
  {Piconcelli}, {Menci}, {Aussel}, \& {Sturm}}]{Cicone12}
{Cicone}, C., {Feruglio}, C., {Maiolino}, R., {et~al.} 2012, \aap, 543, A99

\bibitem[{{Cicone} {et~al.}(2015){Cicone}, {Maiolino}, {Gallerani}, {Neri},
  {Ferrara}, {Sturm}, {Fiore}, {Piconcelli}, \& {Feruglio}}]{Cicone15}
{Cicone}, C., {Maiolino}, R., {Gallerani}, S., {et~al.} 2015, \aap, 574, A14

\bibitem[{{Cicone} {et~al.}(2014){Cicone}, {Maiolino}, {Sturm},
  {Graci{\'a}-Carpio}, {Feruglio}, {Neri}, {Aalto}, {Davies}, {Fiore},
  {Fischer}, {Garc{\'{\i}}a-Burillo}, {Gonz{\'a}lez-Alfonso},
  {Hailey-Dunsheath}, {Piconcelli}, \& {Veilleux}}]{Cicone14}
{Cicone}, C., {Maiolino}, R., {Sturm}, E., {et~al.} 2014, \aap, 562, A21

\bibitem[{{Contursi} {et~al.}(2013){Contursi}, {Poglitsch}, {Gr{\'a}cia
  Carpio}, {Veilleux}, {Sturm}, {Fischer}, {Verma}, {Hailey-Dunsheath}, {Lutz},
  {Davies}, {Gonz{\'a}lez-Alfonso}, {Sternberg}, {Genzel}, \&
  {Tacconi}}]{Contursi13}
{Contursi}, A., {Poglitsch}, A., {Gr{\'a}cia Carpio}, J., {et~al.} 2013, \aap,
  549, A118

\bibitem[{{Croxall} {et~al.}(2017){Croxall}, {Smith}, {Pellegrini}, {Groves},
  {Bolatto}, {Herrera-Camus}, {Sandstrom}, {Draine}, {Wolfire}, {Armus},
  {Boquien}, {Brandl}, {Dale}, {Galametz}, {Hunt}, {Kennicutt}, {Kreckel},
  {Rigopoulou}, {van der Werf}, \& {Wilson}}]{Croxall17}
{Croxall}, K.~V., {Smith}, J.~D., {Pellegrini}, E., {et~al.} 2017, \apj, 845,
  96

\bibitem[{{De Rosa} {et~al.}(2011){De Rosa}, {Decarli}, {Walter}, {Fan},
  {Jiang}, {Kurk}, {Pasquali}, \& {Rix}}]{DeRosa11}
{De Rosa}, G., {Decarli}, R., {Walter}, F., {et~al.} 2011, \apj, 739, 56

\bibitem[{{Decarli} {et~al.}(2014){Decarli}, {Walter}, {Carilli}, {Bertoldi},
  {Cox}, {Ferkinhoff}, {Groves}, {Maiolino}, {Neri}, {Riechers}, \&
  {Weiss}}]{Decarli14}
{Decarli}, R., {Walter}, F., {Carilli}, C., {et~al.} 2014, \apjl, 782, L17

\bibitem[{{Decarli} {et~al.}(2017){Decarli}, {Walter}, {Venemans},
  {Ba{\~n}ados}, {Bertoldi}, {Carilli}, {Fan}, {Farina}, {Mazzucchelli},
  {Riechers}, {Rix}, {Strauss}, {Wang}, \& {Yang}}]{Decarli17}
{Decarli}, R., {Walter}, F., {Venemans}, B.~P., {et~al.} 2017, \nat, 545, 457

\bibitem[{{Decarli} {et~al.}(2018){Decarli}, {Walter}, {Venemans},
  {Ba{\~n}ados}, {Bertoldi}, {Carilli}, {Fan}, {Farina}, {Mazzucchelli},
  {Riechers}, {Rix}, {Strauss}, {Wang}, \& {Yang}}]{Decarli18}
{Decarli}, R., {Walter}, F., {Venemans}, B.~P., {et~al.} 2018, \apj, 854, 97

\bibitem[{{Fabian}(2012)}]{fabian12}
{Fabian}, A.~C. 2012, \araa, 50, 455

\bibitem[{{Fan} {et~al.}(2018){Fan}, {Knudsen}, {Fogasy}, \& {Drouart}}]{Fan18}
{Fan}, L., {Knudsen}, K.~K., {Fogasy}, J., \& {Drouart}, G. 2018, \apjl, 856,
  L5

\bibitem[{{Fan} {et~al.}(2001){Fan}, {Narayanan}, {Lupton}, {Strauss}, {Knapp},
  {Becker}, {White}, {Pentericci}, {Leggett}, {Haiman}, {Gunn}, {Ivezi{\'c}},
  {Schneider}, {Anderson}, {Brinkmann}, {Bahcall}, {Connolly}, {Csabai}, {Doi},
  {Fukugita}, {Geballe}, {Grebel}, {Harbeck}, {Hennessy}, {Lamb}, {Miknaitis},
  {Munn}, {Nichol}, {Okamura}, {Pier}, {Prada}, {Richards}, {Szalay}, \&
  {York}}]{Fan01}
{Fan}, X., {Narayanan}, V.~K., {Lupton}, R.~H., {et~al.} 2001, \aj, 122, 2833

\bibitem[{{Fan} {et~al.}(2006){Fan}, {Strauss}, {Richards}, {Hennawi},
  {Becker}, {White}, {Diamond-Stanic}, {Donley}, {Jiang}, {Kim}, {Vestergaard},
  {Young}, {Gunn}, {Lupton}, {Knapp}, {Schneider}, {Brandt}, {Bahcall},
  {Barentine}, {Brinkmann}, {Brewington}, {Fukugita}, {Harvanek}, {Kleinman},
  {Krzesinski}, {Long}, {Neilsen}, {Nitta}, {Snedden}, \& {Voges}}]{Fan06}
{Fan}, X., {Strauss}, M.~A., {Richards}, G.~T., {et~al.} 2006, \aj, 131, 1203

\bibitem[{{Feruglio} {et~al.}(2017){Feruglio}, {Ferrara}, {Bischetti},
  {Downes}, {Neri}, {Ceccarelli}, {Cicone}, {Fiore}, {Gallerani}, {Maiolino},
  {Menci}, {Piconcelli}, {Vietri}, {Vignali}, \& {Zappacosta}}]{Feruglio17}
{Feruglio}, C., {Ferrara}, A., {Bischetti}, M., {et~al.} 2017, \aap, 608, A30

\bibitem[{{Feruglio} {et~al.}(2015){Feruglio}, {Fiore}, {Carniani},
  {Piconcelli}, {Zappacosta}, {Bongiorno}, {Cicone}, {Maiolino}, {Marconi},
  {Menci}, {Puccetti}, \& {Veilleux}}]{Feruglio15}
{Feruglio}, C., {Fiore}, F., {Carniani}, S., {et~al.} 2015, \aap, 583, A99

\bibitem[{{Gallerani} {et~al.}(2018){Gallerani}, {Pallottini}, {Feruglio},
  {Ferrara}, {Maiolino}, {Vallini}, {Riechers}, \& {Pavesi}}]{Gallerani18}
{Gallerani}, S., {Pallottini}, A., {Feruglio}, C., {et~al.} 2018, \mnras, 473,
  1909

\bibitem[{{Hailey-Dunsheath} {et~al.}(2010){Hailey-Dunsheath}, {Nikola},
  {Stacey}, {Oberst}, {Parshley}, {Benford}, {Staguhn}, \&
  {Tucker}}]{Hailey-Dunsheath10}
{Hailey-Dunsheath}, S., {Nikola}, T., {Stacey}, G.~J., {et~al.} 2010, \apj,
  714, L162

\bibitem[{{Harrison} {et~al.}(2016){Harrison}, {Alexander}, {Mullaney},
  {Stott}, {Swinbank}, {Arumugam}, {Bauer}, {Bower}, {Bunker}, \&
  {Sharples}}]{Harrison16}
{Harrison}, C.~M., {Alexander}, D.~M., {Mullaney}, J.~R., {et~al.} 2016,
  \mnras, 456, 1195

\bibitem[{{Harrison} {et~al.}(2014){Harrison}, {Alexander}, {Mullaney}, \&
  {Swinbank}}]{harrison14}
{Harrison}, C.~M., {Alexander}, D.~M., {Mullaney}, J.~R., \& {Swinbank}, A.~M.
  2014, \mnras, 441, 3306

\bibitem[{{Harrison} {et~al.}(2012){Harrison}, {Alexander}, {Swinbank},
  {Smail}, {Alaghband-Zadeh}, {Bauer}, {Chapman}, {Del Moro}, {Hickox},
  {Ivison}, {Men{\'e}ndez-Delmestre}, {Mullaney}, \& {Nesvadba}}]{Harrison12}
{Harrison}, C.~M., {Alexander}, D.~M., {Swinbank}, A.~M., {et~al.} 2012,
  \mnras, 426, 1073

\bibitem[{{Jameson} {et~al.}(2018){Jameson}, {Bolatto}, {Wolfire}, {Warren},
  {Herrera-Camus}, {Croxall}, {Pellegrini}, {Smith}, {Rubio}, {Indebetouw},
  {Israel}, {Meixner}, {Roman-Duval}, {van Loon}, {Muller}, {Verdugo},
  {Zinnecker}, \& {Okada}}]{Jameson18}
{Jameson}, K.~E., {Bolatto}, A.~D., {Wolfire}, M., {et~al.} 2018, \apj, 853,
  111

\bibitem[{{Janssen} {et~al.}(2016){Janssen}, {Christopher}, {Sturm},
  {Veilleux}, {Contursi}, {Gonz{\'a}lez-Alfonso}, {Fischer}, {Davies}, {Verma},
  {Graci{\'a}-Carpio}, {Genzel}, {Lutz}, {Sternberg}, {Tacconi}, {Burtscher},
  \& {Poglitsch}}]{Janssen16}
{Janssen}, A.~W., {Christopher}, N., {Sturm}, E., {et~al.} 2016, \apj, 822, 43

\bibitem[{{Jiang} {et~al.}(2015){Jiang}, {McGreer}, {Fan}, {Bian}, {Cai},
  {Cl{\'e}ment}, {Wang}, \& {Fan}}]{Jiang15}
{Jiang}, L., {McGreer}, I.~D., {Fan}, X., {et~al.} 2015, \aj, 149, 188

\bibitem[{{Jiang} {et~al.}(2016){Jiang}, {McGreer}, {Fan}, {Strauss},
  {Ba{\~n}ados}, {Becker}, {Bian}, {Farnsworth}, {Shen}, {Wang}, {Wang},
  {Wang}, {White}, {Wu}, {Wu}, {Yang}, \& {Yang}}]{Jiang16}
{Jiang}, L., {McGreer}, I.~D., {Fan}, X., {et~al.} 2016, \apj, 833, 222

\bibitem[{{King} \& {Pounds}(2015)}]{king15}
{King}, A. \& {Pounds}, K. 2015, \araa, 53, 115

\bibitem[{{Knudsen} {et~al.}(2016){Knudsen}, {Richard}, {Kneib}, {Jauzac},
  {Cl{\'e}ment}, {Drouart}, {Egami}, \& {Lindroos}}]{Knudsen16}
{Knudsen}, K.~K., {Richard}, J., {Kneib}, J.-P., {et~al.} 2016, \mnras, 462, L6

\bibitem[{{Kreckel} {et~al.}(2014){Kreckel}, {Armus}, {Groves}, {Lyubenova},
  {D{\'{\i}}az-Santos}, {Schinnerer}, {Appleton}, {Croxall}, {Dale}, {Hunt},
  {Beir{\~a}o}, {Bolatto}, {Calzetti}, {Donovan Meyer}, {Draine}, {Hinz},
  {Kennicutt}, {Meidt}, {Murphy}, {Smith}, {Tabatabaei}, \&
  {Walter}}]{Kreckel14}
{Kreckel}, K., {Armus}, L., {Groves}, B., {et~al.} 2014, \apj, 790, 26

\bibitem[{{Kurk} {et~al.}(2007){Kurk}, {Walter}, {Fan}, {Jiang}, {Riechers},
  {Rix}, {Pentericci}, {Strauss}, {Carilli}, \& {Wagner}}]{Kurk07}
{Kurk}, J.~D., {Walter}, F., {Fan}, X., {et~al.} 2007, \apj, 669, 32

\bibitem[{{Lagache} {et~al.}(2018){Lagache}, {Cousin}, \&
  {Chatzikos}}]{Lagache18}
{Lagache}, G., {Cousin}, M., \& {Chatzikos}, M. 2018, \aap, 609, A130

\bibitem[{{Maiolino} {et~al.}(2005){Maiolino}, {Cox}, {Caselli}, {Beelen},
  {Bertoldi}, {Carilli}, {Kaufman}, {Menten}, {Nagao}, {Omont}, {Wei{\ss}},
  {Walmsley}, \& {Walter}}]{Maiolino05}
{Maiolino}, R., {Cox}, P., {Caselli}, P., {et~al.} 2005, \aap, 440, L51

\bibitem[{{Maiolino} {et~al.}(2012){Maiolino}, {Gallerani}, {Neri}, {Cicone},
  {Ferrara}, {Genzel}, {Lutz}, {Sturm}, {Tacconi}, {Walter}, {Feruglio},
  {Fiore}, \& {Piconcelli}}]{Maiolino12}
{Maiolino}, R., {Gallerani}, S., {Neri}, R., {et~al.} 2012, \mnras, 425, L66

\bibitem[{{Matsuoka} {et~al.}(2016){Matsuoka}, {Onoue}, {Kashikawa}, {Iwasawa},
  {Strauss}, {Nagao}, {Imanishi}, {Niida}, {Toba}, {Akiyama}, {Asami}, {Bosch},
  {Foucaud}, {Furusawa}, {Goto}, {Gunn}, {Harikane}, {Ikeda}, {Kawaguchi},
  {Kikuta}, {Komiyama}, {Lupton}, {Minezaki}, {Miyazaki}, {Morokuma},
  {Murayama}, {Nishizawa}, {Ono}, {Ouchi}, {Price}, {Sameshima}, {Silverman},
  {Sugiyama}, {Tait}, {Takada}, {Takata}, {Tanaka}, {Tang}, \&
  {Utsumi}}]{Matsuoka16}
{Matsuoka}, Y., {Onoue}, M., {Kashikawa}, N., {et~al.} 2016, \apj, 828, 26

\bibitem[{{Mazzucchelli} {et~al.}(2017){Mazzucchelli}, {Ba{\~n}ados},
  {Venemans}, {Decarli}, {Farina}, {Walter}, {Eilers}, {Rix}, {Simcoe},
  {Stern}, {Fan}, {Schlafly}, {De Rosa}, {Hennawi}, {Chambers}, {Greiner},
  {Burgett}, {Draper}, {Kaiser}, {Kudritzki}, {Magnier}, {Metcalfe}, {Waters},
  \& {Wainscoat}}]{Mazzucchelli17}
{Mazzucchelli}, C., {Ba{\~n}ados}, E., {Venemans}, B.~P., {et~al.} 2017, \apj,
  849, 91

\bibitem[{{McMullin} {et~al.}(2007){McMullin}, {Waters}, {Schiebel}, {Young},
  \& {Golap}}]{McMullin07}
{McMullin}, J.~P., {Waters}, B., {Schiebel}, D., {Young}, W., \& {Golap}, K.
  2007, in Astronomical Society of the Pacific Conference Series, Vol. 376,
  Astronomical Data Analysis Software and Systems XVI, ed. R.~A. {Shaw},
  F.~{Hill}, \& D.~J. {Bell}, 127

\bibitem[{{Mortlock} {et~al.}(2009){Mortlock}, {Patel}, {Warren}, {Venemans},
  {McMahon}, {Hewett}, {Simpson}, {Sharp}, {Burningham}, {Dye}, {Ellis},
  {Gonzales-Solares}, \& {Hu{\'e}lamo}}]{Mortlock09}
{Mortlock}, D.~J., {Patel}, M., {Warren}, S.~J., {et~al.} 2009, \aap, 505, 97

\bibitem[{{Mullaney} {et~al.}(2011){Mullaney}, {Alexander}, {Goulding}, \&
  {Hickox}}]{Mullaney11}
{Mullaney}, J.~R., {Alexander}, D.~M., {Goulding}, A.~D., \& {Hickox}, R.~C.
  2011, \mnras, 414, 1082

\bibitem[{{Nanni} {et~al.}(2018){Nanni}, {Gilli}, {Vignali}, {Mignoli},
  {Comastri}, {Vanzella}, {Zamorani}, {Calura}, {Lanzuisi}, {Brusa}, {Tozzi},
  {Iwasawa}, {Cappi}, {Vito}, {Balmaverde}, {Costa}, {Risaliti}, {Paolillo},
  {Prandoni}, {Liuzzo}, {Rosati}, {Chiaberge}, {Caminha}, {Sani}, {Cappelluti},
  \& {Norman}}]{Nanni18}
{Nanni}, R., {Gilli}, R., {Vignali}, C., {et~al.} 2018, \aap, 614, A121

\bibitem[{{Olsen} {et~al.}(2015){Olsen}, {Greve}, {Narayanan}, {Thompson},
  {Toft}, \& {Brinch}}]{Olsen15}
{Olsen}, K.~P., {Greve}, T.~R., {Narayanan}, D., {et~al.} 2015, \apj, 814, 76

\bibitem[{{Omont} {et~al.}(2013){Omont}, {Willott}, {Beelen}, {Bergeron},
  {Orellana}, \& {Delorme}}]{Omont13}
{Omont}, A., {Willott}, C.~J., {Beelen}, A., {et~al.} 2013, \aap, 552, A43

\bibitem[{{Reed} {et~al.}(2015){Reed}, {McMahon}, {Banerji}, {Becker},
  {Gonzalez-Solares}, {Martini}, {Ostrovski}, {Rauch}, {Abbott}, {Abdalla},
  {Allam}, {Benoit-Levy}, {Bertin}, {Buckley-Geer}, {Burke}, {Carnero Rosell},
  {da Costa}, {D'Andrea}, {DePoy}, {Desai}, {Diehl}, {Doel}, {Cunha},
  {Estrada}, {Evrard}, {Fausti Neto}, {Finley}, {Fosalba}, {Frieman}, {Gruen},
  {Honscheid}, {James}, {Kent}, {Kuehn}, {Kuropatkin}, {Lahav}, {Maia},
  {Makler}, {Marshall}, {Merritt}, {Miquel}, {Mohr}, {Nord}, {Ogando},
  {Plazas}, {Romer}, {Roodman}, {Rykoff}, {Sako}, {Sanchez}, {Santiago},
  {Schubnell}, {Sevilla}, {Smith}, {Soares-Santos}, {Suchyta}, {Swanson},
  {Tarle}, {Thomas}, {Tucker}, {Walker}, \& {Wechsler}}]{Reed15}
{Reed}, S.~L., {McMahon}, R.~G., {Banerji}, M., {et~al.} 2015, \mnras, 454,
  3952

\bibitem[{{Roberts-Borsani} \& {Saintonge}(2018)}]{RobertsBorsani18}
{Roberts-Borsani}, G.~W. \& {Saintonge}, A. 2018, ArXiv e-prints
  [\eprint[arXiv]{1807.07575}]

\bibitem[{{Robitaille} \& {Bressert}(2012)}]{Robitaille12}
{Robitaille}, T. \& {Bressert}, E. 2012, {APLpy: Astronomical Plotting Library
  in Python}, Astrophysics Source Code Library

\bibitem[{{Rose} {et~al.}(2018){Rose}, {Tadhunter}, {Ramos Almeida},
  {Rodr{\'{\i}}guez Zaur{\'{\i}}n}, {Santoro}, \& {Spence}}]{Rose18}
{Rose}, M., {Tadhunter}, C., {Ramos Almeida}, C., {et~al.} 2018, \mnras, 474,
  128

\bibitem[{{Runnoe} {et~al.}(2012){Runnoe}, {Brotherton}, \& {Shang}}]{Runnoe12}
{Runnoe}, J.~C., {Brotherton}, M.~S., \& {Shang}, Z. 2012, \mnras, 426, 2677

\bibitem[{{Rupke} \& {Veilleux}(2011)}]{rupke11}
{Rupke}, D.~S.~N. \& {Veilleux}, S. 2011, \apjl, 729, L27

\bibitem[{{Schaye} {et~al.}(2015){Schaye}, {Crain}, {Bower}, {Furlong},
  {Schaller}, {Theuns}, {Dalla Vecchia}, {Frenk}, {McCarthy}, {Helly},
  {Jenkins}, {Rosas-Guevara}, {White}, {Baes}, {Booth}, {Camps}, {Navarro},
  {Qu}, {Rahmati}, {Sawala}, {Thomas}, \& {Trayford}}]{Schaye15}
{Schaye}, J., {Crain}, R.~A., {Bower}, R.~G., {et~al.} 2015, \mnras, 446, 521

\bibitem[{{Sijacki} {et~al.}(2015){Sijacki}, {Vogelsberger}, {Genel},
  {Springel}, {Torrey}, {Snyder}, {Nelson}, \& {Hernquist}}]{sijacki15}
{Sijacki}, D., {Vogelsberger}, M., {Genel}, S., {et~al.} 2015, \mnras, 452, 575

\bibitem[{{Somerville} {et~al.}(2008){Somerville}, {Hopkins}, {Cox},
  {Robertson}, \& {Hernquist}}]{somerville08}
{Somerville}, R.~S., {Hopkins}, P.~F., {Cox}, T.~J., {Robertson}, B.~E., \&
  {Hernquist}, L. 2008, \mnras, 391, 481

\bibitem[{{Spence} {et~al.}(2018){Spence}, {Tadhunter}, {Rose}, \&
  {Rodr{\'{\i}}guez Zaur{\'{\i}}n}}]{Spence18}
{Spence}, R.~A.~W., {Tadhunter}, C.~N., {Rose}, M., \& {Rodr{\'{\i}}guez
  Zaur{\'{\i}}n}, J. 2018, \mnras, 478, 2438

\bibitem[{{Stanley} {et~al.}(2018){Stanley}, {Harrison}, {Alexander},
  {Simpson}, {Knudsen}, {Mullaney}, {Rosario}, \& {Scholtz}}]{Stanley18}
{Stanley}, F., {Harrison}, C.~M., {Alexander}, D.~M., {et~al.} 2018, \mnras,
  478, 3721

\bibitem[{{Sturm} {et~al.}(2011){Sturm}, {Gonz{\'a}lez-Alfonso}, {Veilleux},
  {Fischer}, {Graci{\'a}-Carpio}, {Hailey-Dunsheath}, {Contursi}, {Poglitsch},
  {Sternberg}, {Davies}, {Genzel}, {Lutz}, {Tacconi}, {Verma}, {Maiolino}, \&
  {de Jong}}]{sturm11}
{Sturm}, E., {Gonz{\'a}lez-Alfonso}, E., {Veilleux}, S., {et~al.} 2011, \apjl,
  733, L16

\bibitem[{{Tadhunter} {et~al.}(2018){Tadhunter}, {Rodr{\'{\i}}guez
  Zaur{\'{\i}}n}, {Rose}, {Spence}, {Batcheldor}, {Berg}, {Ramos Almeida},
  {Spoon}, {Sparks}, \& {Chiaber}}]{Tadhunter18}
{Tadhunter}, C., {Rodr{\'{\i}}guez Zaur{\'{\i}}n}, J., {Rose}, M., {et~al.}
  2018, \mnras, 478, 1558

\bibitem[{{Vayner} {et~al.}(2017){Vayner}, {Wright}, {Murray}, {Armus},
  {Larkin}, \& {Mieda}}]{Vayner17}
{Vayner}, A., {Wright}, S.~A., {Murray}, N., {et~al.} 2017, \apj, 851, 126

\bibitem[{{Veilleux} {et~al.}(2017){Veilleux}, {Bolatto}, {Tombesi},
  {Mel{\'e}ndez}, {Sturm}, {Gonz{\'a}lez-Alfonso}, {Fischer}, \&
  {Rupke}}]{Veilleux17}
{Veilleux}, S., {Bolatto}, A., {Tombesi}, F., {et~al.} 2017, \apj, 843, 18

\bibitem[{{Veilleux} {et~al.}(2013){Veilleux}, {Mel{\'e}ndez}, {Sturm},
  {Gracia-Carpio}, {Fischer}, {Gonz{\'a}lez-Alfonso}, {Contursi}, {Lutz},
  {Poglitsch}, {Davies}, {Genzel}, {Tacconi}, {de Jong}, {Sternberg}, {Netzer},
  {Hailey-Dunsheath}, {Verma}, {Rupke}, {Maiolino}, {Teng}, \&
  {Polisensky}}]{Veilleux13}
{Veilleux}, S., {Mel{\'e}ndez}, M., {Sturm}, E., {et~al.} 2013, \apj, 776, 27

\bibitem[{{Venemans}(2017)}]{Venemans17b}
{Venemans}, B.~P. 2017, The Messenger, 169, 48

\bibitem[{{Venemans} {et~al.}(2015){Venemans}, {Ba{\~n}ados}, {Decarli},
  {Farina}, {Walter}, {Chambers}, {Fan}, {Rix}, {Schlafly}, {McMahon},
  {Simcoe}, {Stern}, {Burgett}, {Draper}, {Flewelling}, {Hodapp}, {Kaiser},
  {Magnier}, {Metcalfe}, {Morgan}, {Price}, {Tonry}, {Waters}, {AlSayyad},
  {Banerji}, {Chen}, {Gonz{\'a}lez-Solares}, {Greiner}, {Mazzucchelli},
  {McGreer}, {Miller}, {Reed}, \& {Sullivan}}]{Venemans15}
{Venemans}, B.~P., {Ba{\~n}ados}, E., {Decarli}, R., {et~al.} 2015, \apjl, 801,
  L11

\bibitem[{Venemans {et~al.}(2018)Venemans, Decarli, Walter, Ba{\~{n}}ados,
  Bertoldi, Fan, Farina, Mazzucchelli, Riechers, Rix, Wang, \&
  Yang}]{venemans18}
Venemans, B.~P., Decarli, R., Walter, F., {et~al.} 2018, The Astrophysical
  Journal, 866, 159

\bibitem[{{Venemans} {et~al.}(2012){Venemans}, {McMahon}, {Walter}, {Decarli},
  {Cox}, {Neri}, {Hewett}, {Mortlock}, {Simpson}, \& {Warren}}]{Venemans12}
{Venemans}, B.~P., {McMahon}, R.~G., {Walter}, F., {et~al.} 2012, \apjl, 751,
  L25

\bibitem[{{Venemans} {et~al.}(2017){Venemans}, {Walter}, {Decarli},
  {Ferkinhoff}, {Wei{\ss}}, {Findlay}, {McMahon}, {Sutherland}, \&
  {Meijerink}}]{Venemans17a}
{Venemans}, B.~P., {Walter}, F., {Decarli}, R., {et~al.} 2017, \apj, 845, 154

\bibitem[{{Venemans} {et~al.}(2016){Venemans}, {Walter}, {Zschaechner},
  {Decarli}, {De Rosa}, {Findlay}, {McMahon}, \& {Sutherland}}]{Venemans16}
{Venemans}, B.~P., {Walter}, F., {Zschaechner}, L., {et~al.} 2016, \apj, 816,
  37

\bibitem[{{Wang} {et~al.}(2013){Wang}, {Wagg}, {Carilli}, {Walter}, {Lentati},
  {Fan}, {Riechers}, {Bertoldi}, {Narayanan}, {Strauss}, {Cox}, {Omont},
  {Menten}, {Knudsen}, {Neri}, \& {Jiang}}]{Wang13}
{Wang}, R., {Wagg}, J., {Carilli}, C.~L., {et~al.} 2013, \apj, 773, 44

\bibitem[{{Wang} {et~al.}(2011){Wang}, {Wagg}, {Carilli}, {Walter}, {Riechers},
  {Willott}, {Bertoldi}, {Omont}, {Beelen}, {Cox}, {Strauss}, {Bergeron},
  {Forveille}, {Menten}, \& {Fan}}]{Wang11}
{Wang}, R., {Wagg}, J., {Carilli}, C.~L., {et~al.} 2011, \apjl, 739, L34

\bibitem[{{Wang} {et~al.}(2016){Wang}, {Wu}, {Neri}, {Fan}, {Walter},
  {Carilli}, {Momjian}, {Bertoldi}, {Strauss}, {Li}, {Wang}, {Riechers},
  {Jiang}, {Omont}, {Wagg}, \& {Cox}}]{Wang16}
{Wang}, R., {Wu}, X.-B., {Neri}, R., {et~al.} 2016, \apj, 830, 53

\bibitem[{{Willott} {et~al.}(2017){Willott}, {Bergeron}, \&
  {Omont}}]{willott17}
{Willott}, C.~J., {Bergeron}, J., \& {Omont}, A. 2017, \apj, 850, 108

\bibitem[{{Willott} {et~al.}(2007){Willott}, {Delorme}, {Omont}, {Bergeron},
  {Delfosse}, {Forveille}, {Albert}, {Reyl{\'e}}, {Hill}, {Gully-Santiago},
  {Vinten}, {Crampton}, {Hutchings}, {Schade}, {Simard}, {Sawicki}, {Beelen},
  \& {Cox}}]{Willott07}
{Willott}, C.~J., {Delorme}, P., {Omont}, A., {et~al.} 2007, \aj, 134, 2435

\bibitem[{{Willott} {et~al.}(2010){Willott}, {Delorme}, {Reyl{\'e}}, {Albert},
  {Bergeron}, {Crampton}, {Delfosse}, {Forveille}, {Hutchings}, {McLure},
  {Omont}, \& {Schade}}]{Willott10}
{Willott}, C.~J., {Delorme}, P., {Reyl{\'e}}, C., {et~al.} 2010, \aj, 139, 906

\bibitem[{{Willott} {et~al.}(2013){Willott}, {Omont}, \&
  {Bergeron}}]{Willott13}
{Willott}, C.~J., {Omont}, A., \& {Bergeron}, J. 2013, \apj, 770, 13

\end{thebibliography}

\begin{appendix}
\section{Tests on the spectral stacking methods and orientation effects}\label{appendix}
\begin{figure}[tbh]
\begin{center}

   \includegraphics[width=0.4\textwidth]{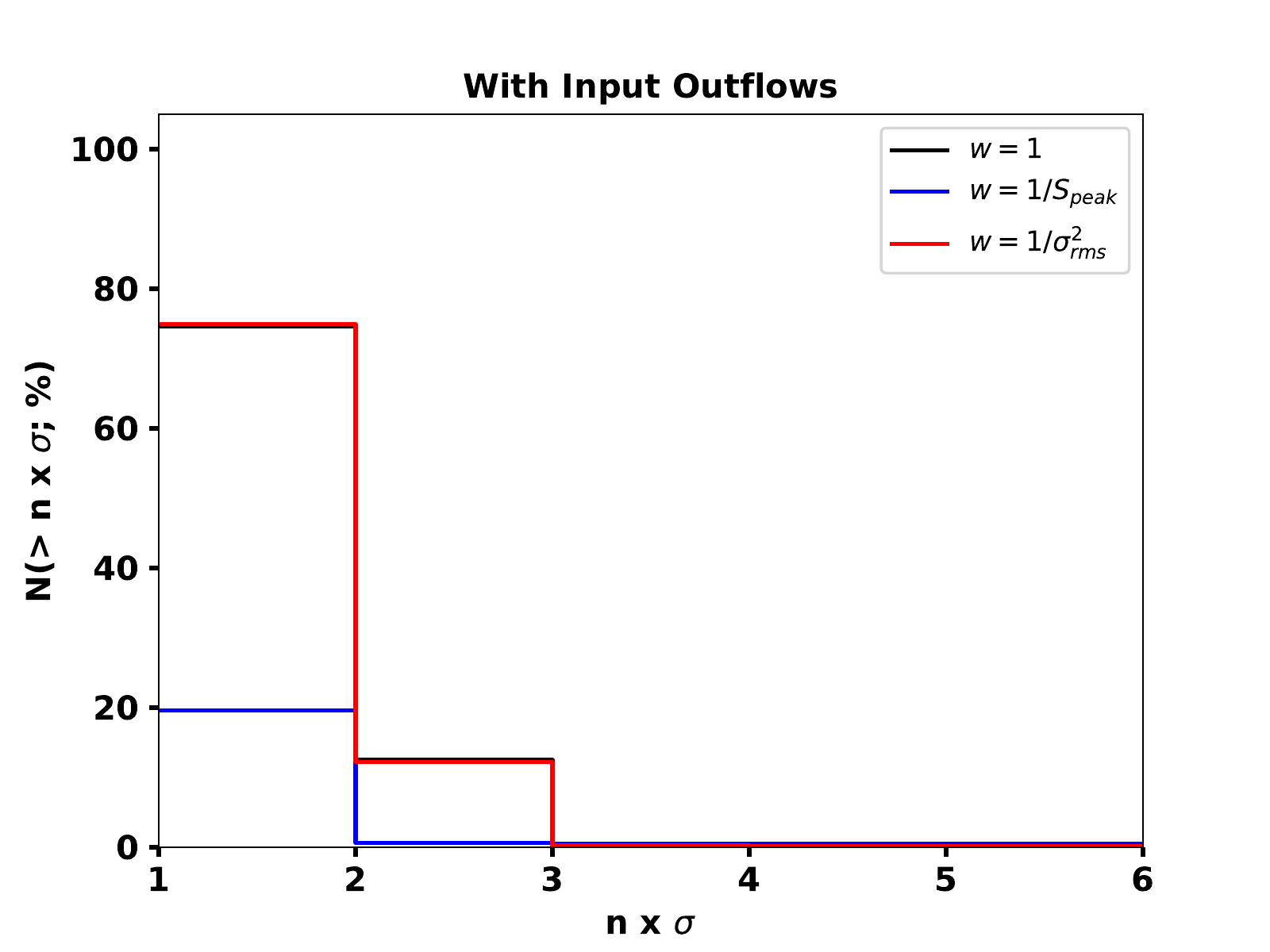}
    \includegraphics[width=0.4\textwidth]{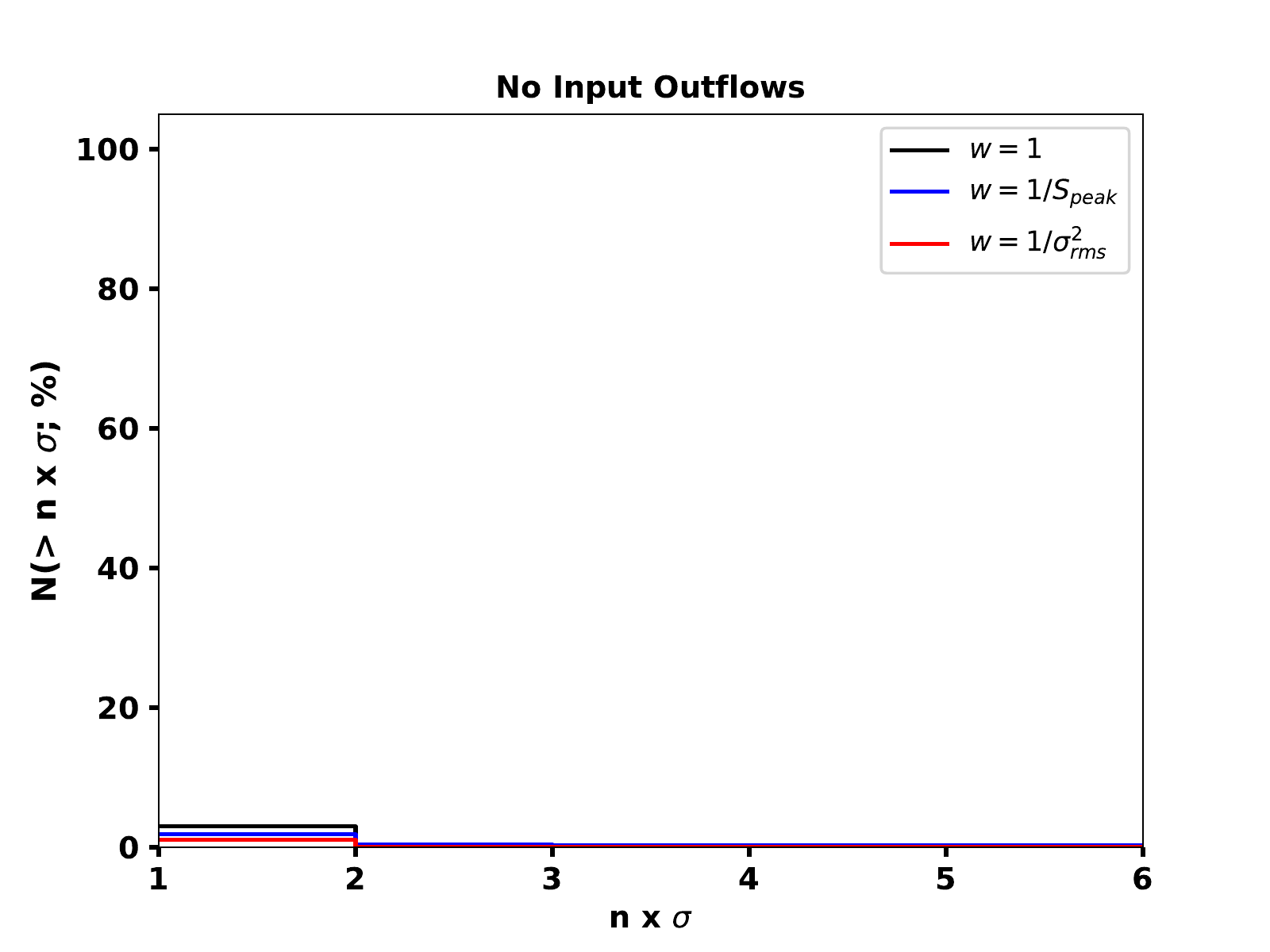}
    
    \caption{Histogram of the fraction of iterations that the stacked spectra shows excess emission at significance levels greater than n\,$\times\,\sigma$. {\em (top)} The results of our test on stacking mock samples where all sources had a broad component below the noise levels. {\em (bottom)} The results of our test on stacking mock samples where all sources had no added broad component. We show the results for stacking with all three of the weights assumed in this work.}
    
    \label{fig:appendix_test1_sigma}
	
\end{center}
\end{figure}

\begin{figure*}
    \centering
    \includegraphics[width=0.98\textwidth]{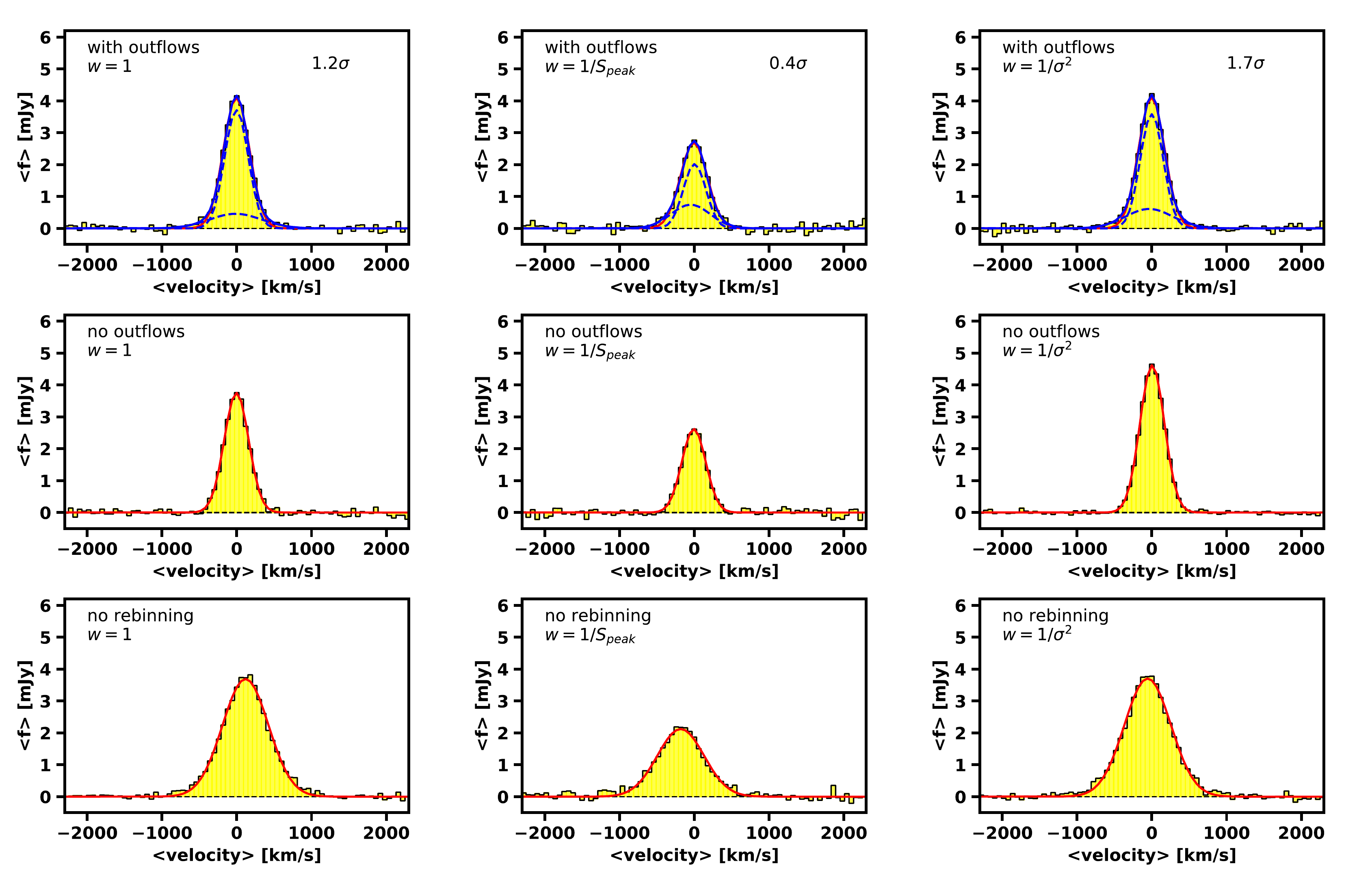}
    \caption{{\em (top)} Example spectra from the test on stacking mock samples where all sources had a broad component below the noise levels, for each of the different weights used. Also given are the significance values of the excess emission due to a broad component, for each example. {\em (middle)} Example spectra from the test on stacking mock samples where all sources had no added broad component. {\em (bottom)} Example spectra from the test on stacking mock samples where all sources had no added broad component, but without using velocity re-binning. We also show the one (red curve) and two-component fits (blue solid and dashed curves) to the stacked lines.}
    \label{fig:appendix_test_spec}
\end{figure*}

In section~5.3 we discussed two tests on the spectral stacking analysis using mock samples. 
One where all the lines in our mock sample included a broad component bellow the noise, 
and one where no broad component was included in the mock sample. 
Here we show the fraction of iterations showing excess emission due to a broad component
at different significance levels (Figure~\ref{fig:appendix_test1_sigma}), and example 
stacked spectra from these tests (Figure~\ref{fig:appendix_test_spec}).
As discussed in section~5.3, our spectral stacking analysis successfully retrieves a broad component
in the stacked lines of the mock samples with an outflow component 75\% of the time with an excess emission of $1-2\sigma$ 
for weights of $w = 1$ and $w = 1/\sigma^2_{rms}$. But for $w = 1/S_{peak}$, its only able to retrieve excess emission at $1-2\sigma$ 20\% of the time. 
In the case of the mock samples without an outflow component included, 
there is no excess emission detected at $>2\sigma$, and only $<17\%$
of the time do we retrieve excess emission between $>0.4\sigma$.

In section~6.1 we discussed the effect of stacking lines without taking into account the range of line widths of the sample, i.e. without applying velocity rebinning on the lines before stacking. In this case we did a test on stacking a mock sample of lines with similar properties to our sample and with no broad component, without applying velocity rebinning. We repeated the exercise with and without the inclusion of noise in the mock samples. If there is no noise included, the stacked line does not have a broad component in any of the iterations.  When including the noise, we retrieve $>2\sigma$ excess emission only 2--5\% of the time.We show examples of the stacked spectra from this test in Fig.~\ref{fig:appendix_test_spec}(last row). 

\begin{figure}
\centering
    \includegraphics[width=0.4\textwidth]{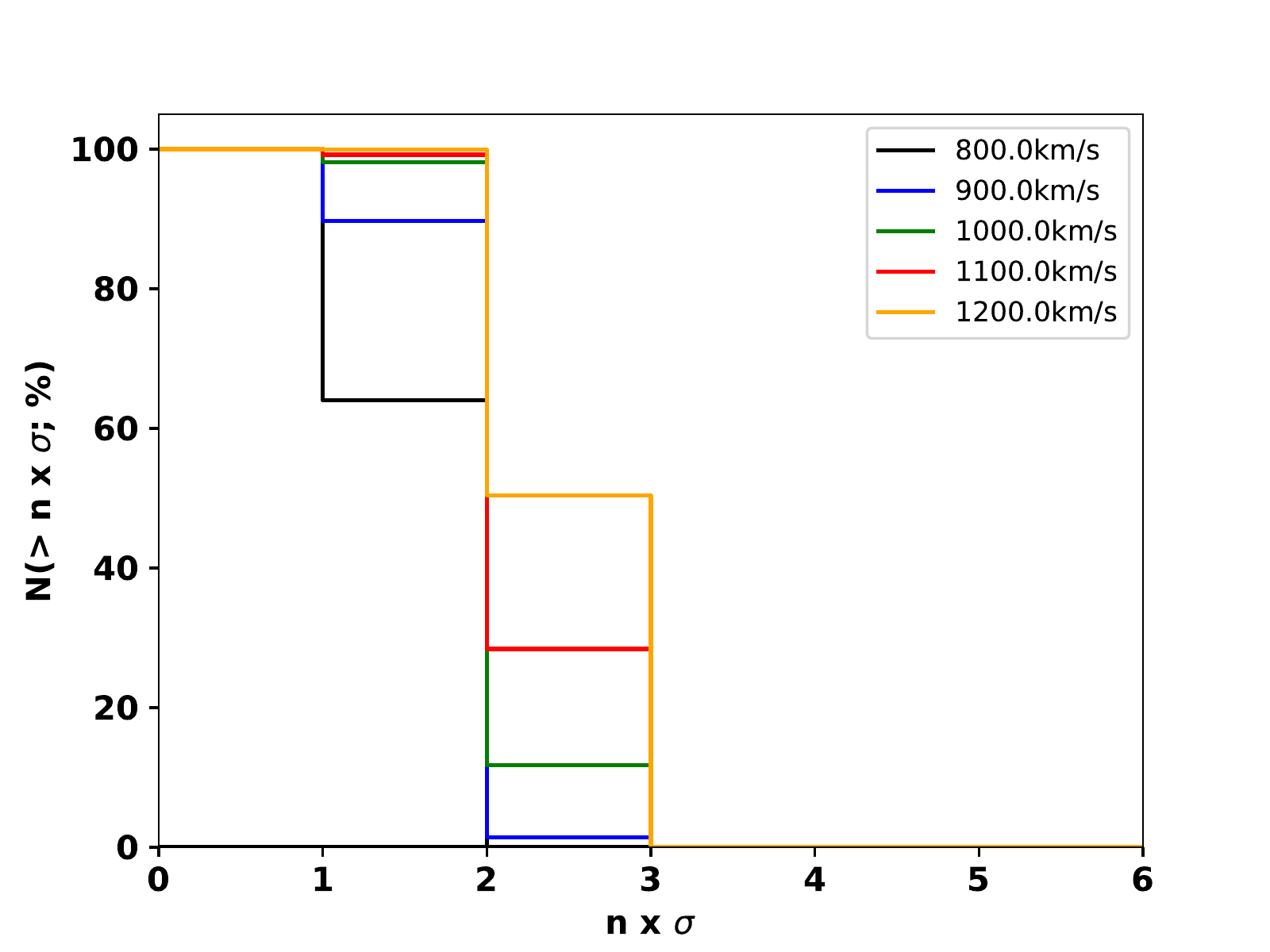}
    \caption{Histogram of the fraction of iterations that the stacked spectra shows excess emission at significance levels greater than n\,$\times\,\sigma$. In this case we show the results of our outflow orientation test. Each colour corresponds to a different initial FWHM of the broad component.}
    \label{fig:appendix_test4_sigma}
\end{figure}

\begin{figure*}
    \centering
    \includegraphics[width=0.98\textwidth]{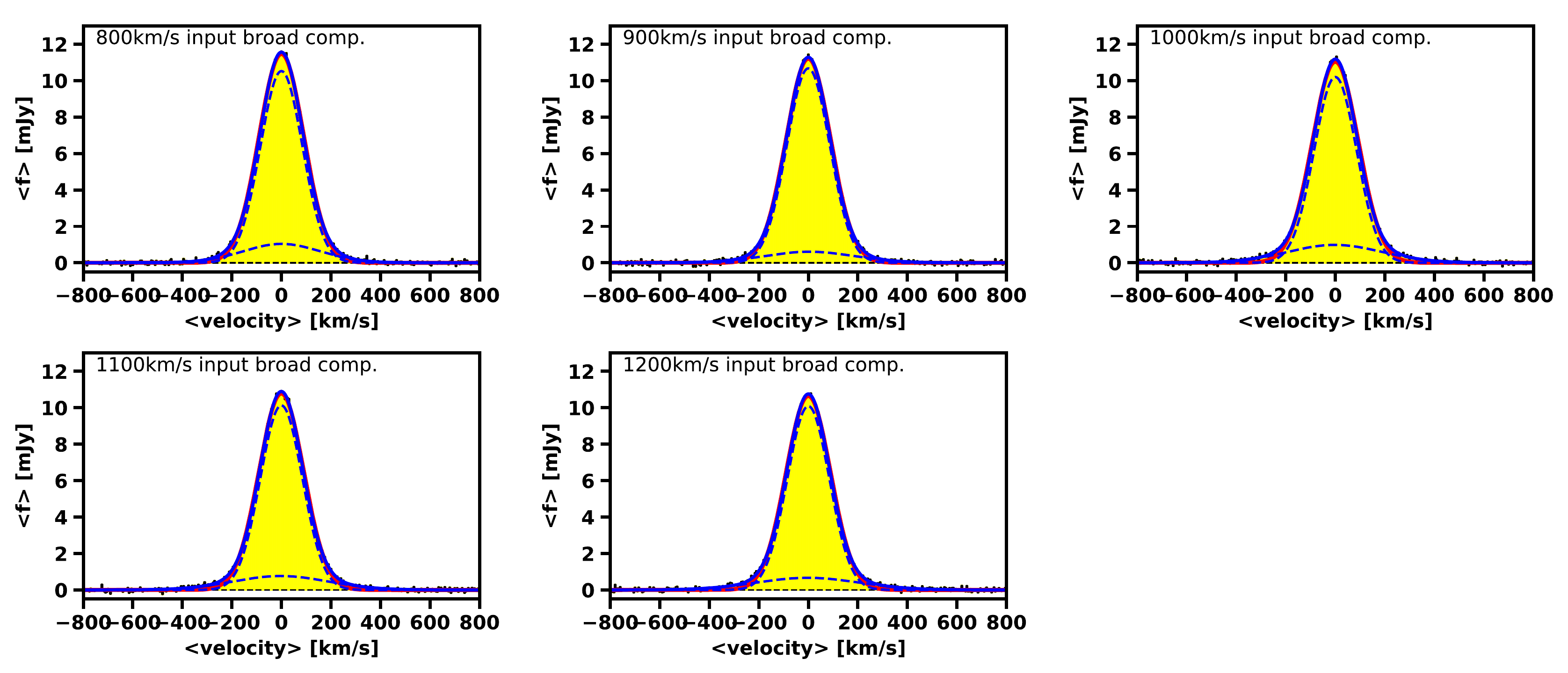}
    \caption{Example spectra from the orientation test, for each of the different initial broad component FWHM values. We also show the one (red curve) and two-component fits (blue solid and dashed curves) to the stacked lines.}
    \label{fig:appendix_orientation_test_spec}
\end{figure*}

In section~6.2.1 we discussed the effect that the orientation of the outflows could have on the detections of a broad component signature in the stacked lines (see section~\ref{sec:orientation}). In Fig.~\ref{fig:appendix_test4_sigma} we show the 
fraction of iterations with excess emission in the stack at different significance levels, for each of the assumed initial FWHM of the broad component (800, 900, 1000, 1100, 1200\,km\,s$^{\rm -1}$). We find that for initial broad components with $\rm FWHM>900$\,km\,s$^{\rm -1}$, we can retrieve excess emission of $>1\sigma$, while we can retrieve a $>2\sigma$ signal only for
initial broad components with $\rm FWHM>1100$\,km\,s$^{\rm -1}$ $\sim$27-50\% of the time. Example spectra from this test are shown in Fig.~\ref{fig:appendix_orientation_test_spec}.

\end{appendix}

\end{document}